\documentclass[trackchanges, twocolumn, twocolappendix]{aastex701}

\usepackage{comment}
\usepackage{multirow}
\usepackage{threeparttable}
\usepackage{url}
\usepackage{natbib}
\usepackage{subcaption}
\usepackage{graphicx}
\usepackage{float}
\usepackage{fancyhdr}
\usepackage{autobreak}
\usepackage{multirow}
\usepackage{xcolor}
\usepackage{adjustbox}
\usepackage{color}

\begin{document}

\title{XRISM Discovery of Multiple Ionized Fe-K Emission and Absorption Components in Centaurus A}

\author[orcid=0000-0002-6960-9274,sname='Taishu Kayanoki']{Taishu Kayanoki}
\affiliation{Department of Physics, Graduate School of Advanced Science and Engineering, Hiroshima University, Hiroshima 739-8526, Japan}
\email[show]{kayanoki@astro.hiroshima-u.ac.jp}

\author[orcid=0000-0002-0921-8837,sname='Yasushi Fukazawa']{Yasushi Fukazawa}
\affiliation{Department of Physics, Graduate School of Advanced Science and Engineering, Hiroshima University, Hiroshima 739-8526, Japan}
\email{}

\author[orcid=0000-0001-7557-9713,sname='Junjie Mao']{Junjie Mao}
\affiliation{Department of Astronomy, Tsinghua University, Beijing 100084, China}
\affiliation{Department of Physics, Graduate School of Advanced Science and Engineering, Hiroshima University, Hiroshima 739-8526, Japan}
\email{}

\author[orcid=0000-0003-2869-7682,sname='Jon M. Miller']{Jon M. Miller}
\affiliation{Department of Astronomy, The University of Michigan, 1085 South University Avenue, Ann Arbor, Michigan, 48103, USA}
\email{}

\author[orcid=0009-0006-4968-7108,sname='Luigi Gallo']{Luigi Gallo}
\affiliation{Saint Mary’s University, Department of Astronomy and Physics, 923 Robie St, Halifax, Nova Scotia B3H 3C3, Canada}
\email{}

\author[sname='Tahir Yaqoob']{Tahir Yaqoob}
\affiliation{National Aeronautics and Space Administration, Goddard Space Flight Center, Greenbelt, MD 20771, USA}
\affiliation{Center for Research and Exploration in Space Science and Technology, NASA/GSFC (CRESST II), Greenbelt, MD 20771, USA}
\affiliation{Center for Space Science and Technology, University of Maryland, Baltimore County (UMBC), 1000 Hilltop Circle, Baltimore, MD 21250, USA}
\email{}

\author[orcid=0000-0002-7962-5446,sname='Richard Mushotzky']{Richard Mushotzky}
\affiliation{Department of Astronomy, University of Maryland, College Park, MD 20742, USA}
\affiliation{Joint Space-Science Institute, University of Maryland, College Park, MD 20742, USA}
\email{}

\author[orcid=0000-0002-5924-4822,sname='David Bogensberger']{David Bogensberger}
\affiliation{Universit\'e Paris-Saclay, Universit\'e Paris Cit\'e, CEA, CNRS, AIM, 91191 Gif-sur-Yvette, France}
\affiliation{Department of Astronomy, The University of Michigan, 1085 South University Avenue, Ann Arbor, Michigan, 48103, USA}
\email{}

\author[orcid=0000-0003-2161-0361,sname='Misaki Mizumoto']{Misaki Mizumoto}
\affiliation{Science Education Research Unit, University of Teacher Education Fukuoka, Akama-bunkyo-machi, Munakata, Fukuoka 811-4192, Japan}
\email{}

\author[orcid=0000-0003-4235-5304,sname='Kouichi Hagino']{Kouichi Hagino}
\affiliation{Department of Physics, The University of Tokyo, 7-3-1 Hongo, Bunkyo-ku, Tokyo 113-0033, Japan}
\email{}

\author[orcid=0000-0001-6020-517X,sname='Hirofumi Noda']{Hirofumi Noda}
\affiliation{Astronomical Institute, Tohoku University, 6-3 Aramakiazaaoba, Aoba-ku, Sendai, Miyagi 980-8578, Japan}
\email{}

\author[orcid=,sname='Yoshihiro Ueda']{Yoshihiro Ueda}
\affiliation{Department of Astronomy, Kyoto University, Kitashirakawa-Oiwake-cho, Sakyo-ku, Kyoto 606-8502, Japan}
\email{}

\author[orcid=0000-0002-5097-1257,sname='Makoto Tashiro']{Makoto Tashiro}
\affiliation{Department of Physics, Saitama University, Saitama 338-8570, Japan}
\affiliation{Institute of Space and Astronautical Science (ISAS), Japan Aerospace Exploration Agency (JAXA), Kanagawa 252-5210, Japan}
\email{}

\author[orcid=0009-0000-9577-8701,sname='Yuya  Nakatani']{Yuya  Nakatani}
\affiliation{Department of Astronomy, Kyoto University, Kitashirakawa-Oiwake-cho, Sakyo-ku, Kyoto 606-8502, Japan}
\email{}

\author[orcid=0000-0002-2304-4773,sname='Toshiya Iwata']{Toshiya Iwata}
\affiliation{Department of Physics, The University of Tokyo, 7-3-1 Hongo, Bunkyo-ku, Tokyo 113-0033, Japan}
\email{}

\author[orcid=0009-0008-3206-235X,sname='Misaki Urata']{Misaki Urata}
\affiliation{Department of Physics, Graduate School of Advanced Science and Engineering, Hiroshima University, Hiroshima 739-8526, Japan}
\email{}


\begin{abstract}
We present the first clear detection of ionized Fe-K emission and absorption components in the nearby radio galaxy Centaurus A, revealed by the high-resolution XRISM/Resolve detector.
In the 6.5–6.9 keV band, XRISM reveals multiple Fe {\sc xxv} and Fe {\sc xxvi} emission components.
One is a broad (with a width of $\sigma = 3000$ km/s) and redshifted ($+$3400 km/s) component, originating at $D = 0.02$ pc from the central black hole.
The other two components are narrow (with a width of $\sigma = 500$ km/s) and exhibit redshifted and blueshifted velocities ($+$2600 km/s and $-$1500 km/s), originating from more distant regions ($D = 0.1$ pc).
The photo-ionized model explains the broader component, while the two narrower components can be explained by either photo-ionization or collisional ionization.
One interpretation is that the broader component is an outflow at $\sim10^2$ $R_{\rm S}$ ($R_{\rm S}$; Schwarzschild radius) and the narrow component is a shock-heated plasma close to the torus, with a possible connection to the JWST-discovered outflow outside the torus.
Two blueshifted absorption lines are detected at $\sim$7.1 keV ($\sim10^4$ km/s) and $\sim$10.6 keV ($\sim10^5$ km/s).
The line significance of the 10.6 keV line is above 98\%.
The absorption line components might be attributed to the broad emission component.
These results demonstrate the high potential of XRISM/Resolve to characterize ionized emission and absorption features in the Fe-K band.
Our findings establish a new benchmark in the study of circumnuclear environments in low-luminosity radio galaxies, thereby contributing to a broader understanding of AGN unification.
\end{abstract}



\section{Introduction}
Active Galactic Nuclei (AGN) emit radiation across a wide wavelength range, from radio to gamma-rays. 
In non-beamed AGNs, the X-ray emission is primarily a power-law component from the hot corona close to the central supermassive black hole, and it is often absorbed by the dust torus in the case of type-2 AGNs. 
Additionally, a reflection component from the cold dust torus is associated with the neutral Fe-K emission line \citep[e.g.,][]{Pounds1989, Awaki1991, Yaqoob2004, Shu2010, Fukazawa2011}. 
Moreover, ionized emission lines in the Fe-K band have been detected in several AGNs, such as NGC 7314 \citep{Yaqoob2003}, M81 \citep{Page2004}, Mrk 590 \citep{Longinotti2007}, IRAS 18325$-$5926 \citep{Lobban2014}, etc. 
However, due to the poor energy resolution in the CCD spectra and the smaller effective area of grating spectra, it has been difficult to detect ionized emission lines in the Fe-K band of AGNs.
Thus, it remains difficult to determine whether such ionized emission lines truly exist, whether they are a common feature of AGNs, and to distinguish whether the ionization mechanism is photo-ionization or collisional ionization.

AGNs accrete mass onto supermassive black holes (SMBHs) and often exhibit outflows as relativistic jets or ionized winds. 
Ionized winds are characterized and classified by their ionization parameter $\xi$, hydrogen column density $N_{\rm H}$, and outflow velocity. 
The ionization parameter is defined as 
\begin{eqnarray}\label{eqr1}
\xi = {L_{\rm ion} \over n_{\rm H} r^2}
\end{eqnarray}
where $L_{\rm ion}$ ($1-1000$ Ryd, 1 Ryd = 13.6 eV) denotes the ionizing source luminosity, $n_{\rm H}$ represents the hydrogen number density of the disk wind, and $r$ indicates the distance from the source \citep{Tarter1969, Gallo2023}. 
Ultrafast outflows \citep[UFOs,][]{Tombesi2010} with velocities faster than $10^4$ km s$^{-1}$ can be very energetic \citep{Xrism2025} and are highly ionized with ionization parameters $\log \xi~({\rm erg~cm~s^{-1}})=2-7$ and high column densities of $\log N_{\rm H}~({\rm cm^{-2}})=22-25$ \citep{Laha2021, Yamada2024}. 
They are generally observed as blueshifted absorption lines at $7-10$ keV \citep{Tombesi2010}. 
Based on the CCD data prior to XRISM, their occurrence rate is estimated to be $\sim40\%$ of X-ray bright AGN \citep{Tombesi2010, Tombesi2011, Gofford2013}, but is poorly determined.

AGNs are classified as either radio-loud or radio-quiet based on their radio intensity. 
The spectral energy distributions (SEDs) of radio-loud and radio-quiet AGNs are broadly similar across the infrared to X-ray bands; however, their radio and gamma-ray luminosities differ owing to the strong relativistic jets present in the radio-loud AGNs \citep[e.g.][]{Urry1995, Kayanoki2022}. 
In principle, radio galaxies are radio-loud AGNs with jets misaligned from the line of sight, which allows for the observation of emissions from both the jets and the central core in such galaxies.
Therefore, radio galaxies are ideal objects for investigating the relationship between the jet and the accretion disk. 
Recently, several groups have conducted sample studies on radio galaxies, focusing on topics such as UFOs \citep[e.g.,][]{Tombesi2014}, X-ray neutral absorption \citep{Kayanoki2022}, and GeV gamma-ray emission \citep{Fukazawa2016, Fukazawa2022, Matake2023}.

Centaurus A (Cen A), the nearest radio galaxy at a distance of $3.8 \pm 0.1$ Mpc \citep{Harris2010}, has an optical spectrum dominated by narrow lines, so it is classified as a type-2 AGN \citep{Beckmann2011} or LINER with a black hole mass of $(5.5 \pm 0.3) \times 10^7 M_\odot$ \citep{Neumayer2007, Cappellari2009, Koss2022}. 
The Schwarzschild radius is $R_{\rm S} = 2GM_{\rm BH} / c^2 = 1.63\times10^{13}~{\rm cm}$. 
Based on its radio lobe morphology, Cen A is classified as a Fanaroff–Riley type I galaxy \citep{Fanaroff1974}. 
In the soft X-ray band, an emission line component is present from the hot plasma of its host galaxy \citep[e.g.,][Urata et al. in prep.]{Tombesi2014, Fukazawa2016}. 
\citet{Tombesi2014} analyzed the {\it Suzaku} and {\it XMM-Newton} CCD spectra deeply and detected an absorption line in the Fe-K band.
This feature is most likely Fe {\sc xxv}, and the best-fit parameters of an ionized absorption model yielded $\log\xi\sim4.3$ erg cm s$^{-1}$, $N_{\rm H}\sim4\times10^{22}$ cm$^{-2}$, and $v_{\rm out}<0.005c$. 
\citet{Alonso2025} reported results from  JWST/MIRI observations of the Cen A core, identifying an ionized flowing component.
This component consists of a narrow ($\sigma\sim600$ km s$^{-1}$) and fast (reaching approximately $+$1000, $-$1400 km s$^{-1}$) flow in the inner 6 pc region. 
However, no significant ionized emission component in the Fe-K band had been previously detected in past grating spectra \citep{Bogensberger2024}.

The first XRISM paper on Cen A \citep{Bogensberger2025} focused on an analysis of the neutral Fe K$\alpha$ fluorescence line. 
In this paper, we focus on the ionized emission line and the absorption line features using broadband XRISM/Resolve data and simultaneous NuSTAR spectra obtained almost simultaneously. 

\section{Observation and data reduction}
Cen A was observed by XRISM \citep{Tashiro2025} in August 2024 (Table \ref{cenaDATA}) with simultaneous observations by {\it NuSTAR} and {\it Chandra} (Table \ref{cenaDATA}). 
In this analysis, we used the XRISM/Resolve \citep{Ishisaki2025,kelley2025} data for the line study and the {\it NuSTAR} spectrum \citep{Harrison2013} to determine the hard X-ray continuum. 

\begin{center}
\begin{table*}[htbp]
 \caption{Cen A observations}
 \label{cenaDATA}
 \small\centering
  \begin{threeparttable}
  \begin{tabular}{llccc}
\noalign{\smallskip}\hline\hline\noalign{\smallskip}
Satellite & Obs\_ID & Observation start (UT) & Observation end (UT) & Net Exp (ksec)\\
\hline\noalign{\smallskip}
{\it XRISM} & 300019010 & 2024/08/04 09:17:04 & 2024/08/09 16:24:04 & 225.1 \\\noalign{\smallskip}
{\it NuSTAR} & 60901034002 & 2024/08/05 07:01:09 & 2024/08/05 19:26:09 & 22.1 \\\noalign{\smallskip}
{\it NuSTAR} & 60901034004 & 2024/08/09 02:16:09 & 2024/08/09 19:21:09 & 72.4 \\\noalign{\smallskip}
{\it Chandra} & 29490 & 2024/08/07 06:51:52 & 2024/08/07 13:21:36 & 20.4 \\\noalign{\smallskip}
\hline\noalign{\smallskip}
\end{tabular}
\begin{tablenotes}
\item
\end{tablenotes}
\end{threeparttable}
\end{table*}
\end{center}

\subsection{XRISM}
The data reduction for XRISM/Resolve is the same as that described in \citet{Bogensberger2025}. 
The data from the XRISM pipeline were reprocessed using the XRISM software in HEASoft release version 6.34 and the XRISM CalDB version 9 of the XRISM calibration database. 
The large matrices were used for the spectral response matrices (RMFs). 
The ancillary response file (ARF) was created for a single attitude bin, as described in the Quick-Start Guide, version 2.1.

The flux level of Cen A in the XRISM/Resolve spectrum is at least 20 times higher than that of the estimated NXB background. 
Therefore, we did not subtract the background spectrum, similar to the analysis of  NGC 4151 \citep{XRISM2024ngc}. 
X-ray emission from Cen A is somewhat contaminated by X-ray radiation from the jet, X-ray binaries, and diffuse X-ray emission, which are not resolved by XRISM. 
From the {\it Chandra} simultaneous observations, the flux of the X-ray binaries and jet in Cen A is estimated to be less than $10\%$ of the Resolve spectrum, and thus their contribution was ignored in our analysis.

\subsection{NuSTAR}
\begin{figure}[htb]
 \centering
 \includegraphics[width=\hsize]{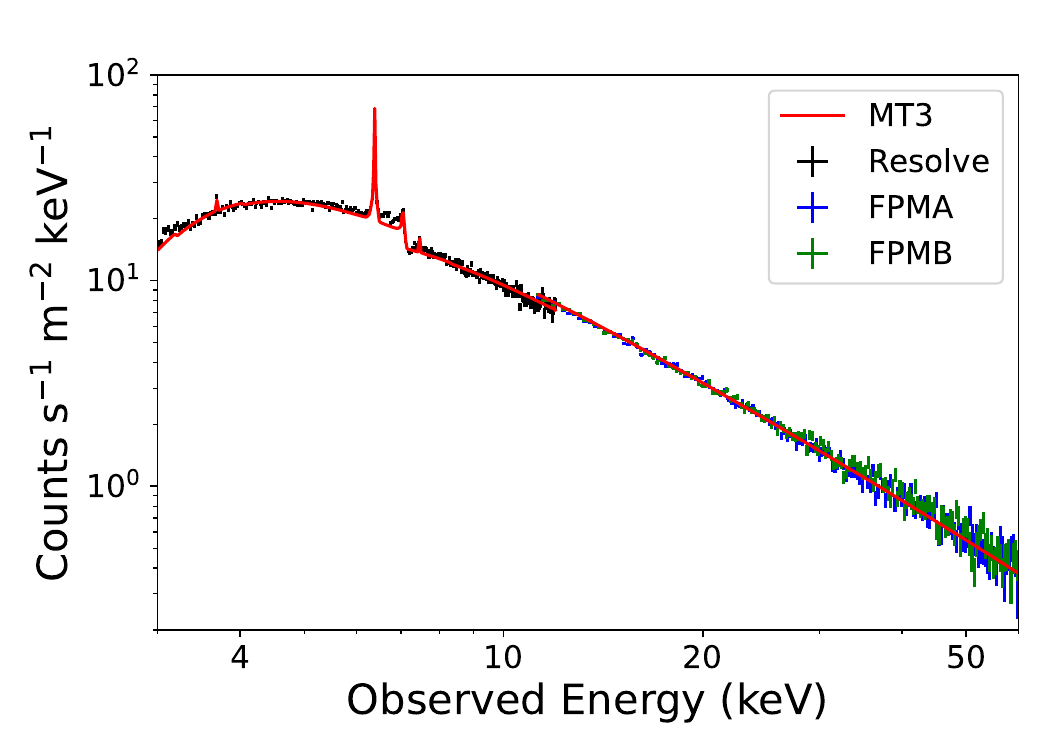}
 \caption{XRISM/Resolve ($3-12$ keV) and NuSTAR ($11.2-60$ keV) spectra with the model curve of the MT3 model, wherein the black point, the blue point, the green point, and the red line represent the Resolve spectra, FPMA spectra, FPMB spectra, and the model curve of the MT3 model, respectively. The Resolve spectrum is binned by a factor of 10 after the optimal binning for visual clarity.}
 \label{MT3}
\end{figure}

The {\it NuSTAR} data were reprocessed using the CalDB updated on September 30, 2024, which is applicable to the Cen A observation period. 
In the {\it NuSTAR} observation, there was a solar flare; therefore, the time intervals with background count rates $>3\sigma$ were filtered out with the Chandra Interactive Analysis of Observations \citep[CIAO,][]{Fruscione2006} version 4.16 {\tt lc\_sigma\_clip} task. 
The flux level and photon index of the FPMA and FPMB spectra for each observation ID were consistent within a $1\sigma$ error. 
We then combined the two observations using the HEASOFT (ver. 6.34) {\tt addspec} tool. 
However, we did not combine the FPMA and FPMB spectra. 
The flux level and photon index of the combined spectra were also consistent with those of the non-combined spectra.  
To correct for cross-calibration between the XRISM/Resolve and NuSTAR spectra, the FPMA and FPMB spectra were rescaled by a factor of 1.08, based on the flux level in the $11-12$ keV energy range (Fig. \ref{MT3}).

\section{Spectral analysis}
We performed the spectral analysis using the SPEX code v3.08.01 \citep{Kaastra2024}. 
For the spectral analysis, the XRISM/Resolve spectra were used in the energy range of $3.0-12.0$ keV, and the {\tt obin} command of the SPEX code was applied to perform optimal binning, following \citet{Kaastra2016}. 
The two NuSTAR spectra (FPMA \& FPMB) were utilized for the $11.2-60.0$ keV range, and we also applied the optimal binning. 
The spectral fitting was performed on the XRISM/Resolve and NuSTAR spectra using the {\it C}-statistic \citep{Kaastra2017}.
To evaluate the fitting results, we used the corrected Akaike information criterion \citep[AICs,][]{Akaike1974, Sugiura1978, Liddle2007}. 
Following \citet{Buhariwalla2024}, we consider $\Delta$AIC values better than 6 to indicate that one model is significantly preferred over another at the 95\% confidence level \citep{Tan2012}. 
In all fitting models, two neutral absorption models ({\tt hot} model in SPEX) were multiplied to account for the absorption by our Galaxy and the host galaxy of Cen A.
The column density for our Galaxy was fixed to $N_{\rm H} = 2.36\times10^{20}$ ${\rm cm}^{-2}$ \citep{HI4PI}, while that of the Cen A host galaxy was fitted as a free parameter at the cosmological redshift.
For reflection, absorption, and emission models, we consider the protosolar abundances of \citet{Lodders2009}. 
For the redshift, we adopt $z = 0.001877$ \citep{Koss2022} for the spectral fitting. 
The errors for each parameter represent the $1\sigma$ confidence level.

\subsection{Continuum modeling}
Before analyzing the lines, we performed continuum modeling of the $3-60$ keV X-ray spectra (Fig. \ref{MT3}). 
First, we fitted the spectra with an absorbed power law. 
We refer to this model as the Absorbed Power-law model (AP model), which is defined as 
\begin{eqnarray}\label{mpmodel}
{\tt hot\times red\times (hot \times etau\times pow) }
\end{eqnarray}
where {\tt reds} is the redshift model. 
The cut-off energy for the power-law component was fixed at 370 keV \citep{Ricci2018} using the {\tt etau} model in SPEX. 
Our results are independent of the cut-off energy.
The photon index becomes $\Gamma = 1.926\pm0.004$, and the hydrogen column density in the host galaxy becomes $N_{\rm H} = (1.68\pm0.01)\times10^{23}$ cm$^{-2}$. 
This model provides a good fit to the continuum well in this energy band.
Hereafter, we used this model as a baseline continuum model.

\subsection{Search for lines}
\begin{figure}
 \centering
 \begin{minipage}[b]{\hsize}
 \centering
 \includegraphics[width=\hsize]{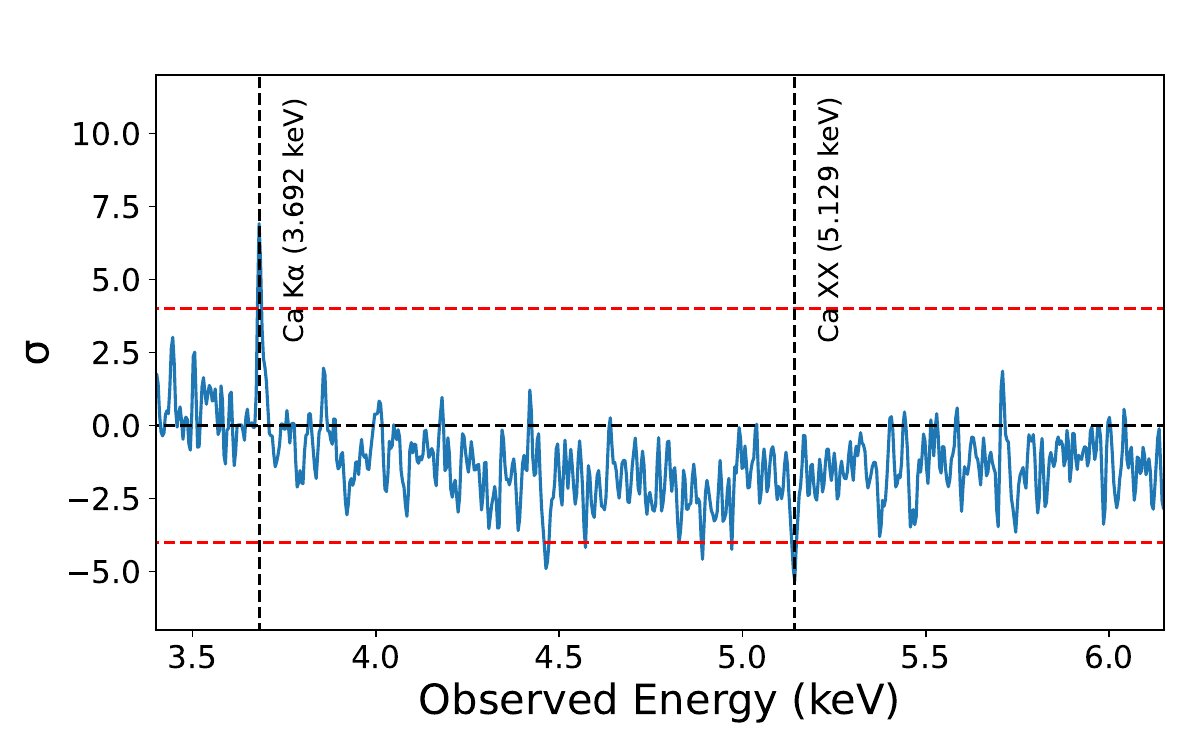}
 \subcaption{$3.4-6.1$ keV band}
\end{minipage}\\
 \begin{minipage}[b]{\hsize}
 \centering
 \includegraphics[width=\hsize]{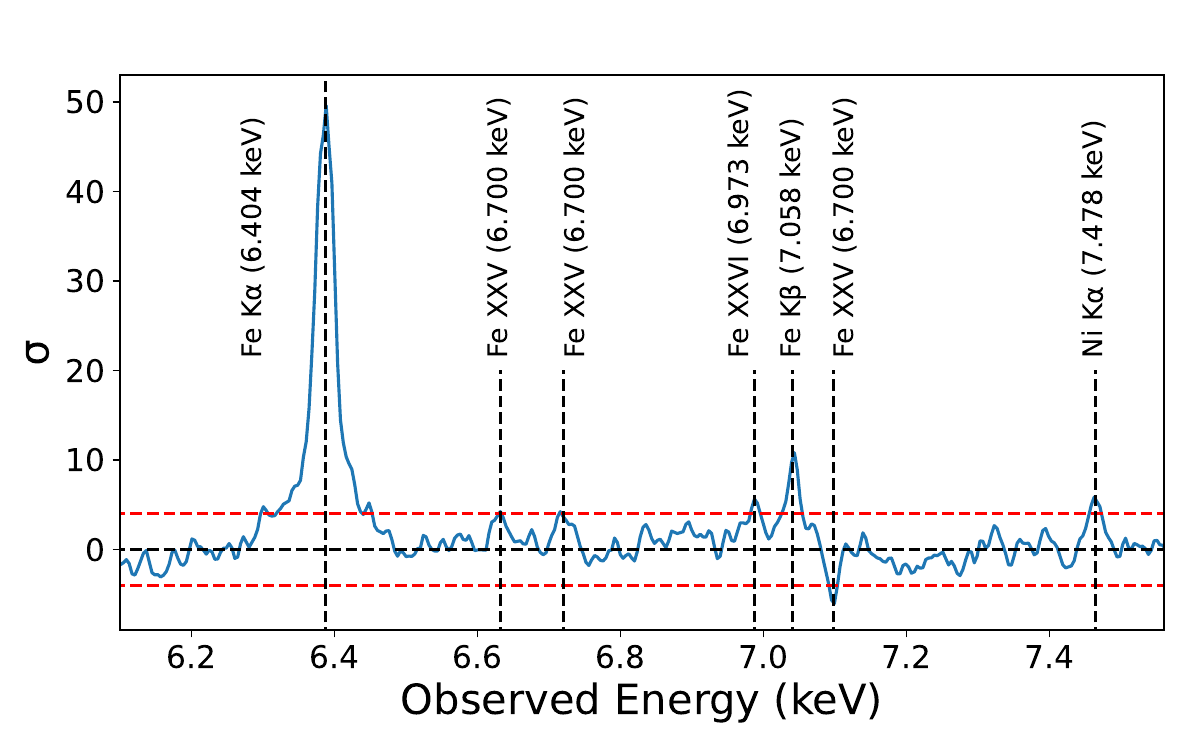}
 \subcaption{$6.1-7.5$ keV band}
\end{minipage}\\
  \begin{minipage}[b]{\hsize}
 \centering
 \includegraphics[width=\hsize]{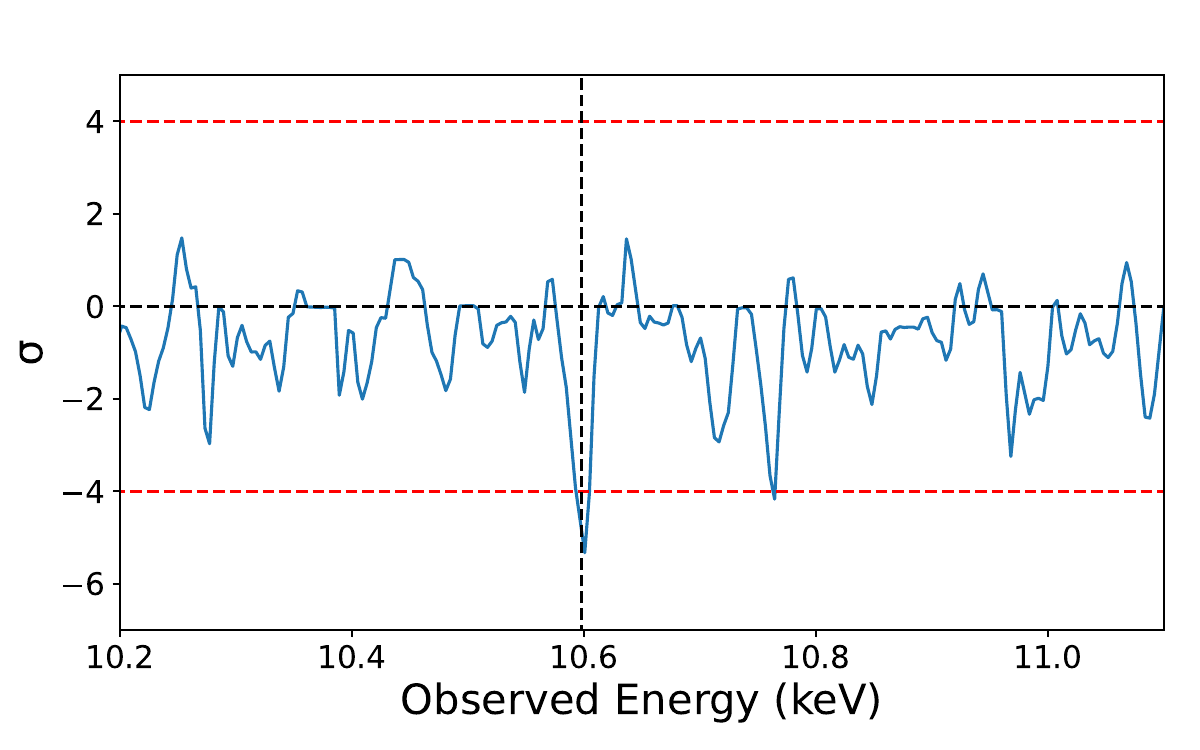}
  \subcaption{$10.2-11.1$ keV band}
  \end{minipage}
 \caption{Results of sliding Gaussian for line search. The step is 4 eV, and the Gaussian FWHM was fixed at 8 eV. Horizontal red dashed lines represent the 4 $\sigma$ confidence level. }
 \label{slidFig1}
\end{figure}

Based on the AP model, we performed the line search using a sliding Gaussian technique. 
The Gaussian FWHM was fixed at 8 eV, roughly twice the Resolve resolution, and this search is optimized for features with $\sim400$ km s$^{-1}$ width at 6 keV. 
The line center energy was varied from 3.5 to 12.0 keV in steps of 4 eV. 
Fig. \ref{slidFig1} shows the significance of the line against the search energy. 
Line candidates above the 4 $\sigma$ confidence level were again fitted by letting the FWHM and line energy vary freely. 
The results are summarized in Table \ref{line4sig}. 
When the narrow emission lines could not be fit due to the presence of broadened emission line components, the FWHM was fixed at 30 eV. 
Significance levels at $4-6$ keV may not be properly evaluated due to residuals (Fig. \ref{slidFig1}a). 
The cause of the residuals is that the AP model was used, and the Fe-K emission lines are not explained. 
However, since this energy band is not the focus of this paper, it does not affect the results and discussion of the ionization component.
As a result, seven emission lines and three absorption lines were found in the Resolve spectrum above the $4\sigma$ confidence level. 
Two additional cases with a step of 1 eV and a Gaussian FWHM of 2 eV or a step of 2 eV and a Gaussian FWHM of 4 eV were also investigated, and it was confirmed that these two cases gave the same result as the case of a step of 4 eV and a Gaussian FWHM of 8 eV.

Considering that the number of trials is ($(12.0-3.5)/0.004=2125$), the line significance would still be over 90\%, for all lines in the table. 
For example, in this trial, the significance of the 10.6 keV absorption line is $4.61\sigma$ (99.99980\%). 
Therefore, the post-trial significance is considered to be $1 - (1-0.9999980031)\times2125=0.9957565$ ($2.63\sigma$). 
In addition, to evaluate the significance of the 10.6 keV absorption line feature, we ran Monte Carlo simulations using the {\tt simulate} command in the SPEX code to quantify the incidence of fake lines when sliding a Gaussian between 7.5 and 12.0 keV. 
We simulated 1,000 times using the same net exposure time. 
To check the probability of detecting a Gaussian absorption trough due to random fluctuations in the simulated data, we simulated with the power-law model and performed the line search by adding a Gaussian function and sliding it.
The Gaussian FWHM was fixed at 8 eV, and the line center energy was changed with a step of 4 eV increments from 7.5 to 12.0 keV. 
The number of the absorption lines above the 4 $\sigma$ confidence level was 104 in the 1,000 simulations. 
The absorption line candidates above the 4 $\sigma$ confidence level were again fitted by letting the FWHM and line energy vary freely. 
Then, the number of absorption lines above the 4 $\sigma$ confidence level was 41 in the 1,000 simulations. 
However, 23 of the 41 lines only covered 1-2 bins, and the obtained FWHM was almost 0 eV. 
From Table \ref{line4sig}, the FWHM of the observed lines is over 10 eV. 
Thus, after excluding the 23 liens with 0 eV FWHM, we found that $18/1000\sim1.8$\% of the simulated spectra produce spurious lines.

Most of the lines in the line search are considered Fe-K lines, but the fluorescent Ca-K line, Ni-K line, and a 5.15 keV absorption line are also present. 
The latter might be a blueshifted Ca-K line. 
Its velocity is similar to that of the ionized Fe-K emission lines. 
These weak ionized emission and absorption lines would not be resolved with CCD resolution, because they would be buried in the tail of the strong neutral Fe-K$\alpha$ line. 
In this paper, we focus on the ionized Fe-K emission and absorption lines. 
The neutral Ca-K and Ni-K will be treated in a future paper.

\begin{table*}[htbp]
 \caption{Lines above the 4 $\sigma$ confidence level}
 \label{line4sig}
 \centering
  \begin{threeparttable}
  \begin{tabular}{clcccccc}
\noalign{\smallskip}\hline\hline\noalign{\smallskip}
$E_{\rm obs}$ ${\rm ^a}$ (keV) & ID & $E_{\rm rest}$ ${\rm ^b}$ (keV) & $\sigma$ ${\rm ^c}$ & FWHM (eV) & $v_{\rm shift}$ ${\rm ^d}$ (km/s) & Emit/Abs ${\rm ^e}$\\
\noalign{\smallskip}\hline\noalign{\smallskip}
$3.6885\pm0.0007$ & Ca K$\alpha$ & $3.6917$  & $5.8$ & $6\pm2$ & $260\pm60$ & E \\\noalign{\smallskip}
$5.151\pm0.003$ & Ca {\sc xx} & $5.129$ & $4.3$ & $22^{+12}_{-6}$ & $-1300\pm200$ & A \\\noalign{\smallskip}
$6.3989^{+0.0002}_{-0.0003}$ & Fe K$\alpha$ & $6.4038$ & $59.7$ & $29.7\pm0.7$ & $220\pm10$ & E\\\noalign{\smallskip}
$6.644^{+0.003}_{-0.004}$ & Fe {\sc xxv} & $6.700$ & $5.9$ & $30$ (fix) & $2500^{+100}_{-200}$ & E \\\noalign{\smallskip}
$6.733\pm0.003$ & Fe {\sc xxv} & $6.700$ & $5.2$ & $30^{+7}_{-6}$ & $-1400\pm100$ & E \\\noalign{\smallskip}
$7.000\pm0.003$ & Fe {\sc xxvi} & $6.973$ & $7.2$ & $30$ (fix) & $-1100\pm100$ & E \\\noalign{\smallskip}
$7.054^{+0.001}_{-0.002}$ & Fe K$\beta$ & $7.058$ & $12.7$ & $30$ (fix) & $170^{+40}_{-80}$ & E \\\noalign{\smallskip}
$7.111\pm0.001$ & Fe {\sc xxv} & $6.700$ & $4.4$ & $9^{+4}_{-3}$ & $-12810\pm40$ & A \\\noalign{\smallskip}
$7.478\pm0.002$ & Ni K$\alpha$ & $7.478$ & $5.8$ & $21^{+5}_{-4}$ & $0\pm80$ & E \\\noalign{\smallskip}
$10.618\pm0.002$ & Fe {\sc xxv} & $6.700$ & $5.5$ & $16\pm5$ & $-129120\pm50$ & A \\\noalign{\smallskip}
\hline
\end{tabular}
\begin{tablenotes}
\item
a: Observed line center energy in Cen A frame. 
b: Theoretical Energy (rest-frame). The line identifications are derived from the AtomDB database 3.1.3 \citep{atomdb2025}. 
c: Significance level of the line. 
d: velocity shift. 
e: "E" means an emission line, and "A" means an absorption line. 
\end{tablenotes}
\end{threeparttable}
\end{table*}

\subsection{Physical modeling for the neutral fluorescence lines}
As a next step, we modeled the neutral fluorescence lines.
Following \citet{Bogensberger2025}, these lines were modeled with the {\tt MyTorus} model \citep{Murphy2009}. 
The model file is {\tt mytl\_V000HLZnEp000\_v01.fits} file \citep{Yaqoob2024} was used as the {\tt MyTorus} model. 
For modeling the line broadening due to Doppler and relativistic effects, the {\tt spei} model \citep{Speith1995} was applied to the {\tt MyTorus} model, corresponding to the {\tt rdblur} model in XSPEC as used in \citet{Bogensberger2025}.  
The photon index and normalization of the {\tt MyTorus} model were linked to the power law ({\tt pow}) model. 
The inclination angle, inner radius, and outer radius of the {\tt spei} model were fixed to those of Model C in \citet{Bogensberger2025} to save computing time. 
As the Model C in \citet{Bogensberger2025}, three {\tt MyTorus} models were used to represent the neutral Fe-K lines. 
In our fitting, the free parameters were the three column densities of {\tt MyTorus}. 
Because the Ni-K$\alpha$ and the Ca-K$\alpha$ lines are not included in the {\tt MyTorus} model, the Gaussian models were added to express them.
In summary, the model to fit the neutral fluorescence lines is 
\begin{eqnarray}\label{mt3model}
{\tt hot\times reds\times \{ hot \times pow + \sum_{i=1}^{2}gaus_i}\nonumber\\
{\tt + \sum_{i=1}^{3}(MyTorus_i \times pow \times spei_i)\} }
\end{eqnarray}
Hereafter, we refer to this model as the MT3 model. 
Fig. \ref{MT3} shows the fitting result around the neutral Fe-K lines, and Table \ref{MT3param} summarizes the best-fit parameters.

\begin{figure*}[htb]
 \centering
 \includegraphics[width=0.9\hsize]{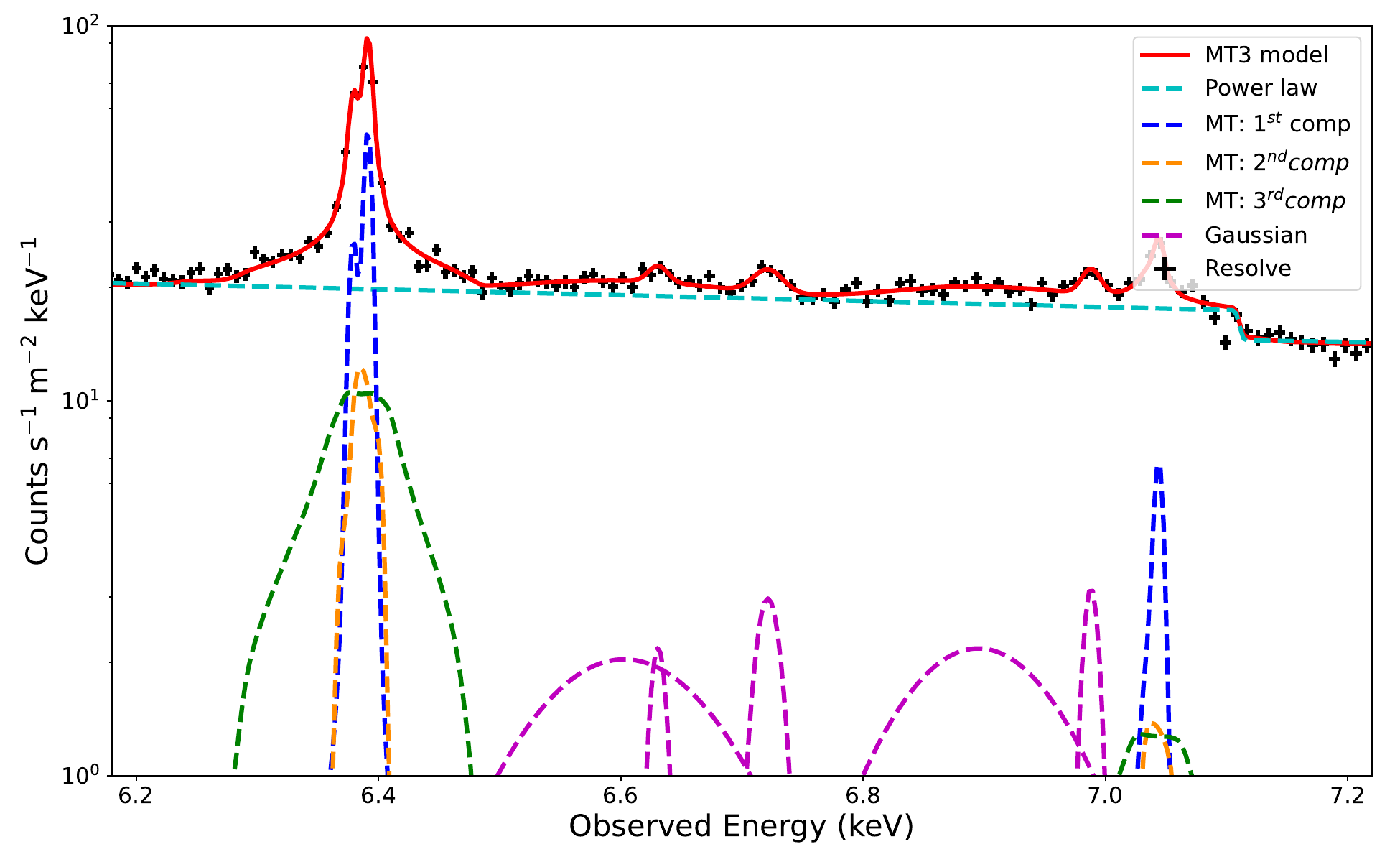}
 \caption{XRISM/Resolve ($6.2-7.2$ keV) spectra with the model curve of the MT3 plus five Gaussian model, wherein the black point, the red line, the cyan dashed-line, the blue dashed-line, the orange dashed-line, the green dashed-line, and the magenta dashed-line represent the Resolve spectra, MT3 model, power-low component, first, second, third {\tt MyTorus} component, and Gaussian component, respectively. The Resolve spectrum is binned by a factor of 3 after the optimal binning for visual clarity. }
 \label{MT_6keV}
\end{figure*}

Parameter values are almost consistent with those of \citet{Bogensberger2025}. 
The small difference is attributed to the different energy band used for fitting. 
Looking at Fig. \ref{MT_6keV}, the neutral Fe-K line profiles are well represented. 
However, significant residuals are seen between the Fe-K$\alpha$ and Fe-K$\beta$ lines, and they are not explained by a single or a few narrow lines. 
Such a residual was also reported in \citet{Bogensberger2025}, suggesting the existence of ionized Fe-K emission lines.

\begin{table}[htbp]
 \caption{Best-fit parameters of the MT3 model. }
 \label{MT3param}
 \small\centering
  \begin{threeparttable}
  \begin{tabular}{lcc}
\noalign{\smallskip}\hline\hline\noalign{\smallskip}
Parameter & Unit & MT3 model\\
\noalign{\smallskip}\hline\noalign{\smallskip}
{\tt pow} $\Gamma$ $^{(1)}$ &  & $1.876\pm0.004$ \\\noalign{\smallskip}
{\tt pow} Norm $^{(2)}$ &  & $6.80\pm0.07$ \\\noalign{\smallskip}
{\tt hot} $N_{\rm H}$ $^{(3)}$ & $10^{23}$ cm$^{-2}$ & $1.56\pm0.01$ \\\noalign{\smallskip}
\hline\noalign{\smallskip}
{\tt MyTorus} $N_{\rm H}$ $^{(4)}$ & $10^{23}$ cm$^{-2}$ & $1.12^{+0.08}_{-0.07}$ \\\noalign{\smallskip}
{\tt spei} $R_{\rm in}$ $^{(6)}$ & $r_{\rm g}$ & $7.2\times10^5$ (fixed) \\\noalign{\smallskip}
{\tt spei} $R_{\rm out}$ $^{(7)}$ & $r_{\rm g}$ & $7.9\times10^5$ (fixed) \\\noalign{\smallskip}
{\tt spei} $i$ $^{(8)}$ & deg & $30$ (fixed) \\\noalign{\smallskip}
{\tt spei} $q$ $^{(9)}$ &  & $2.2$ (fixed) \\\noalign{\smallskip}
\hline\noalign{\smallskip}
{\tt MyTorus} $N_{\rm H}$ $^{(4)}$ & $10^{23}$ cm$^{-2}$ & $0.44\pm0.08$ \\\noalign{\smallskip}
{\tt spei} $R_{\rm in}$ $^{(6)}$ & $r_{\rm g}$ & $5.7\times10^4$ (fixed) \\\noalign{\smallskip}
{\tt spei} $R_{\rm out}$ $^{(7)}$ & $r_{\rm g}$ & $6.3\times10^{4}$ (fixed) \\\noalign{\smallskip}
\hline\noalign{\smallskip}
{\tt MyTorus} $N_{\rm H}$ $^{(4)}$ & $10^{23}$ cm$^{-2}$ & $1.6\pm0.1$ \\\noalign{\smallskip}
{\tt spei} $R_{\rm in}$ $^{(6)}$ & $r_{\rm g}$ & $1.0\times10^3$ (fixed) \\\noalign{\smallskip}
{\tt spei} $R_{\rm out}$ $^{(7)}$ & $r_{\rm g}$ & $3.3\times10^4$ (fixed) \\\noalign{\smallskip}
\hline\noalign{\smallskip}
$C$-stat / d.o.f. &  & 4033.8 / 3360 \\\noalign{\smallskip}
AICs &  & 4058.0 \\\noalign{\smallskip}
\hline
\end{tabular}
\begin{tablenotes}
\item
(1) Photon index of the power-law component. 
(2) Normalization of the power-law component in $10^{50} ~{\rm ph~s^{-1}~keV^{-1}}$ at 1 keV. 
(3) Column density of neutral ISM gas in the host galaxy of the AGN. 
(4) Column density of the reflector. 
(5) Normalization of the {\tt MyTorus} in $10^{50} ~{\rm photons~s^{-1}~keV^{-1}}$ at 1 keV. This value is coupled to the normalization of the X-ray power-law component. 
(6) Inner radius of the reflector. 
(7) Outer radius of the reflector. 
(8) Inclination angle of the reflector (angle between the line of sight and the rotation axis of the reflector). 
(9) Index of radial emissivity power-law profile of the reflector. 
\end{tablenotes}
\end{threeparttable}
\end{table}

\subsection{Gaussian modeling for ionized Fe-K emission lines}
As Fig. \ref{MT_6keV} shows, the measured spectrum exceeds the best-fit MT3 between 6.5-7.0 keV. 
First, we tried to use Gaussian models to represent this excess. 
Table \ref{line4sig} suggested that at least three Gaussian model components are needed. 
Then we introduced three Gaussian model components and thawed the FWHM, line energy, and normalization parameters of the Gaussian model. 
Their FWHM were $\sim70$ eV (for redshifted Fe {\sc xxv}), $\sim30$ eV (for blueshifted Fe {\sc xxv}), and $\sim70$ eV (for redshifted Fe {\sc xxvi}). 
The Gaussian model component for blueshifted Fe {\sc xxv} was consistent with the result in Table \ref{line4sig}. 
However, even with this model, narrow line features still remained at 6.64 keV and 6.98 keV.
Therefore, we added two more Gaussian model components (a total of five for the $6.5-7.0$ keV excess) to represent these features (Fig. \ref{MT_6keV}). 
These lines are likely a redshifted Fe {\sc xxv} line and a blueshifted Fe {\sc xxvi} line. 
As a result, three ionized components were required to represent the excess at $6.5-7.0$ keV. 


\subsection{Physical modeling of the ionized emission line features}

In this subsection, we will fit the Fe-K band excess using physical models.
We applied for the photo-ionization (PIE) model \citep[{\tt pion} model,][]{Gesu2017, Mao2018} and the collisional ionization (CIE) model \citep[{\tt cie} model,][]{Kaastra1996}.

\subsubsection{Photo-ionized emission modeling}
The {\tt pion} model calculates the transmission and emission of a slab of photo-ionized plasma based on the spectrum of the ionizing source, which can be estimated from the broadband fitting along with detailed emission and absorption features \citep{Mao2019}. 
The free parameters were a hydrogen column density $N_{\rm H}$ (cm$^{-2}$), ionization parameter $\log \xi$ (erg cm s$^{-1}$), and outflow velocity $v_{\rm out}$ (km s$^{-1}$). 
Here, the observed Cen A power-law component without any absorption was assumed as the spectrum of the ionizing source. 
In addition, there are two covering factors in the {\tt pion} model. 
The covering factor of the absorber was fixed to $f_{\rm cov}=0$ for emission modeling. 
The covering factor of emission ($\Omega / 4\pi$) was initially left free,
but only a lower limit was obtained as $\Omega /4\pi>0.17$ for all {\tt pion} emission components. 
This is because the covering factor and column density are degenerate \citep{Gesu2017}. 
Therefore, we fixed the covering factor for emission to $\Omega /4\pi=0.2$ for all the {\tt pion} components.

For Doppler broadening, we adopted the {\tt vgau} model to the {\tt pion} emission component. 
The turbulence velocity ($v_{\rm turb}$) of the {\tt pion} emission component and the broadening velocity ($v_{\rm broad}$) cannot be distinguished because these parameters are degenerate \citep{Mao2018}. 
Therefore,  $v_{\rm turb}$ was fixed to be 100 km s$^{-1}$ (default SPEX value) and $v_{\rm broad}$ of the {\tt vgau} model was left free. 
In summary, the free parameters for this fitting were column density ($N_{\rm H}$), ionization parameter ($\log\xi$), Doppler shift velocity ($v_{\rm shift}$), and broadening velocity ($v_{\rm broad}$).

Although three ionized emission components were suggested, we first applied two pion models to represent the Fe-K band excess. 
As a result, the Doppler shift velocity were 3400 km s$^{-1}$ and $-$1500 km s$^{-1}$ with the broadening velocity of 3000 km s$^{-1}$ and 600 km s$^{-1}$, respectively.
The $C$-statistics and AICs of this model were 3643.6 and 3683.9 for the degree of freedom of 3350. 
Although the fit improved by adding two {\tt pion} models compared to the MT3 model ($\Delta$AIC $=-374.1$), this model cannot explain the narrow redshifted component. 
Therefore, we added one more {\tt pion} model component to explain the narrow/redshifted component. 
As a result, the statistics improved significantly ($\Delta$AIC $=-8.2$, Table \ref{emitPIONTable}). 
The final model is
\begin{eqnarray}\label{mt3e2model}
{\tt hot\times reds\times \{hot \times pow + \sum_{i=1}^{2}gaus_i}\nonumber\\
{\tt + \sum_{i=1}^{3}(MyTorus_i \times pow \times spei_i) }\nonumber\\
{\tt + \sum_{i=1}^{3}(pion_i\times pow \times vgau_i)\} }
\end{eqnarray}
We refer to this as the Narrow PION $+$ Broad PION model (Table \ref{emitPIONTable}). 
The results of the Narrow PION $+$ Broad PION model fitting are shown in Fig. \ref{CIE}, and the best-fit parameters are summarized in Table \ref{emitPIONTable}. 

The Doppler shift velocity of the three {\tt pion} emission components are $2600\pm100$ km s$^{-1}$ (Em1), $-1400\pm100$ km s$^{-1}$ (Em2) and $4600^{+700}_{-800}$ km s$^{-1}$ (Em3), and the broadening velocity of three {\tt pion} emission components become $400^{+200}_{-100}$ km s$^{-1}$, $600^{+200}_{-100}$ km s$^{-1}$ and $3300^{+700}_{-500}$ km s$^{-1}$ for Em1, Em2 and Em3, respectively. 
Interestingly, the broad component (Em3) is redshifted with $4600^{+700}_{-800}$ km s$^{-1}$, indicating the ionized material is flowing.
This is the first time that the ionized emission line in Cen A has been detected significantly, and the large broadening velocity indicates that Em3 is located in the inner region rather than the torus. 
We will discuss these features in Section \ref{discuss}.

\begin{table}[htbp]
\caption{Best-fit parameters of the Narrow PION $+$ Broad PION model. }
\label{emitPIONTable}
\small\centering
\begin{threeparttable}
\begin{tabular}{lcc}
\noalign{\smallskip}\hline\hline\noalign{\smallskip}
\multirow{2}{*}{Parameter} & \multirow{2}{*}{Unit} & Narrow PION\\
 &  & $+$ Broad PION model\\
\noalign{\smallskip}\hline\noalign{\smallskip}
\multicolumn{3}{c}{Component Em1}\\\noalign{\smallskip}
{\tt pion} $N_{\rm H}$ $^{(1)}$ & $10^{22}$ cm$^{-2}$ & $2.8^{+1.0}_{-0.9}$ \\\noalign{\smallskip}
{\tt pion} $\log \xi$ $^{(2)}$ & erg cm s$^{-1}$ & $3.10^{+0.07}_{-0.08}$ \\\noalign{\smallskip}
{\tt pion} $v_{\rm shift}$ $^{(3)}$ & km s$^{-1}$ & $2600\pm100$ \\\noalign{\smallskip}
{\tt vgau} $v_{\rm broad}$ $^{(4)}$ & km s$^{-1}$ & $400^{+200}_{-100}$ \\\noalign{\smallskip}
\hline\noalign{\smallskip}
\multicolumn{3}{c}{Component Em2}\\\noalign{\smallskip}
{\tt pion} $N_{\rm H}$ $^{(1)}$ & $10^{22}$ cm$^{-2}$ & $6\pm1$ \\\noalign{\smallskip}
{\tt pion} $\log \xi$ $^{(2)}$ & erg cm s$^{-1}$ & $3.14\pm0.04$ \\\noalign{\smallskip}
{\tt pion} $v_{\rm shift}$ $^{(3)}$ & km s$^{-1}$ & $-1400\pm100$ \\\noalign{\smallskip}
{\tt vgau} $v_{\rm broad}$ $^{(4)}$ & km s$^{-1}$ & $600^{+200}_{-100}$ \\\noalign{\smallskip}
\hline\noalign{\smallskip}
\multicolumn{3}{c}{Component Em3}\\\noalign{\smallskip}
{\tt pion} $N_{\rm H}$ $^{(1)}$ & $10^{23}$ cm$^{-2}$ & $1.7^{+0.6}_{-0.4}$ \\\noalign{\smallskip}
{\tt pion} $\log \xi$ $^{(2)}$ & erg cm s$^{-1}$ & $3.20^{+0.05}_{-0.04}$ \\\noalign{\smallskip}
{\tt pion} $v_{\rm shift}$ $^{(3)}$ & km s$^{-1}$ & $4600^{+700}_{-800}$ \\\noalign{\smallskip}
{\tt vgau} $v_{\rm broad}$ $^{(4)}$ & km s$^{-1}$ & $3300^{+700}_{-500}$ \\\noalign{\smallskip}
\hline\noalign{\smallskip}
$C$-stat / d.o.f. &  & 3627.3 / 3346 \\\noalign{\smallskip}
AICs &  & 3675.7 \\\noalign{\smallskip}
\hline
\end{tabular}
\begin{tablenotes}
\item 
(1) Column density of the {\tt pion} component. 
(2) Logarithm of the ionization parameter of the {\tt pion} component. 
(3) Flowing velocity of the {\tt pion} component, a positive value means the inflow and a negative value means the outflow. 
(4) Broadening velocity of the {\tt vgau} component. 
The turbulence velocity of the {\tt pion} component was fixed to $100$ km s$^{-1}$, and the covering factor for emission was fixed to $0.2$. 
\end{tablenotes}
\end{threeparttable}
\end{table}

\subsubsection{Collisional ionized emission modeling}\label{cieFit}
Next, we replaced the {\tt pion} model with the {\tt cie} model \citep{Kaastra1996}. 
The {\tt cie} model describes a thermal emission from a hot gas in collisional ionization equilibrium without a thermal Doppler broadening. 
For the {\tt cie} modeling, the free parameters were normalization ($Y$) and the electron temperature. 
The normalization measures $Y\equiv n_{\rm H} n_{\rm e}V$, where $n_{\rm H}$ and $n_{\rm e}$ are the electron and hydrogen densities and $V$ is the volume of the source. 
In addition, we adopted the {\tt vgau} model to account for thermal and other broadening effects and the {\tt reds} model for the bulk Doppler velocity shift of the {\tt cie} model. 
In summary, the model for fitting the ionized Fe-K emission lines is
\begin{eqnarray}\label{mt3cie2model}
{\tt hot\times reds\times \{hot \times pow + \sum_{i=1}^{2}gaus_i}\nonumber\\
{\tt + \sum_{i=1}^{3}(MyTorus_i \times pow \times spei_i) }\nonumber\\
{\tt + \sum_{i=1}^{3}(cie\,_i\times reds\,_i \times vgau_i)\} }
\end{eqnarray}
We refer to this as the Narrow CIE $+$ Broad CIE model.

\begin{figure*}[htb]
 \centering
 \includegraphics[width=\hsize]{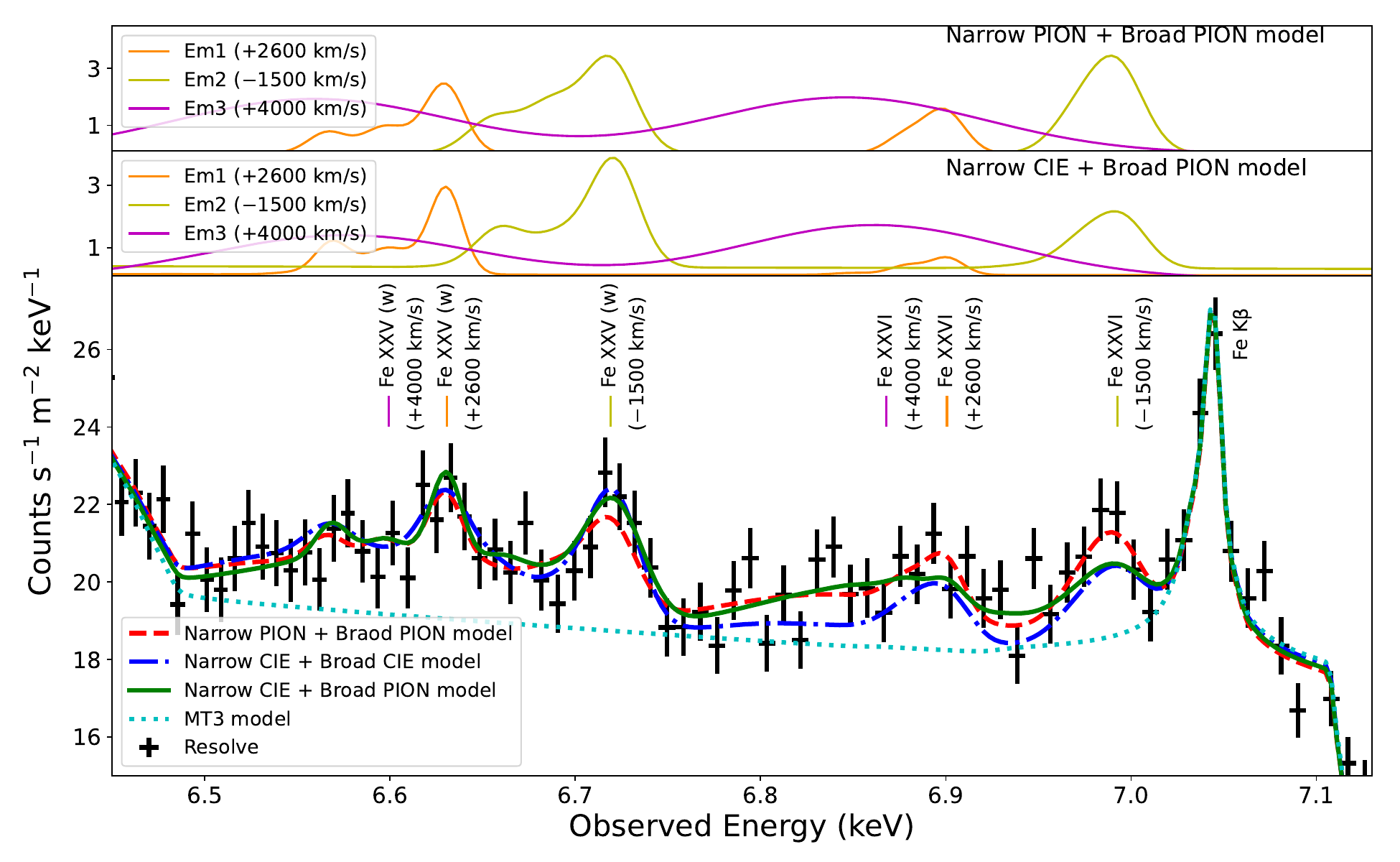}
 \caption{XRISM/Resolve ($6.5-7.1$ keV) spectra with the ionized emission model curve. The vertical and horizontal axes are the same in the top, middle, and bottom figures. Top figure: the model component curves of the Narrow PION $+$ Broad PION model, wherein the orange line, the yellow line, and the magenta line represent the Em1 component (PIE), the Em2 component (PIE), and the Em3 component (PIE), respectively. Middle figure: the model component curves of the Narrow CIE $+$ Broad PION model, wherein the orange line, the yellow line, and the magenta line represent the Em1 component (CIE), the Em2 component (CIE), and the Em3 component (PIE), respectively. Bottom figure: XRISM/Resolve spectra with the model curve of the Narrow PION $+$ Broad PION model, Narrow CIE $+$ Broad CIE model, Narrow CIE $+$ Broad PION model, and MT3 model, wherein the black point, the red dashed line, the blue dot-dashed line, the green line, and the cyan dot line represent the Resolve spectra, Narrow PION $+$ Broad PION model, Narrow CIE $+$ Broad CIE model, Narrow CIE $+$ Broad PION model, and MT3 model, respectively. The Resolve spectrum is binned by a factor of 3 after the optimal binning for visual clarity.}
 \label{CIE}
\end{figure*}

The fitting result of the Narrow CIE $+$ Broad CIE model is shown in Fig. \ref{CIE}, and the best-fit parameters are summarized in Table \ref{CIETable}.
The broadening velocity and the Doppler shift velocity are consistent with those of the Narrow PION $+$ Broad PION model. 
Compared with the Narrow PION $+$ Broad PION model, the Narrow CIE $+$ Broad CIE model could not adequately explain the Fe {\sc xxvi} lines (especially for the broader component, Fig. \ref{CIE}). 
The observed Fe {\sc xxvi} line is stronger than the Fe {\sc xxv} line for the broad component, which requires a temperature above 15 keV. 
On the other hand, the Fe-K emission becomes too weak to reproduce the observed line intensity for such a high temperature.
Consequently, the C-statistics and AIC favor the Narrow PION $+$ Broad PION model ($\Delta$AIC $=+25.8$). 
Thus, we considered the Narrow PION $+$ Broad PION model to be more suitable for the $6.5-7.0$ keV excess than the Narrow CIE $+$ Broad CIE model. 
The broadening velocity of 550 km s$^{-1}$ for the narrow component is larger than the thermal broadening of Fe ions ($\sigma\sim50$ km s$^{-1}$) for the best-fit temperature of 7.5 keV, indicating the presence of other broadening effects of 500 km s$^{-1}$.

\begin{table}[htbp]
 \caption{Best-fit parameters of the Narrow CIE $+$ Broad CIE model. }
 \label{CIETable}
 \small\centering
  \begin{threeparttable}
  \begin{tabular}{lcc}
\noalign{\smallskip}\hline\hline\noalign{\smallskip}
\multirow{2}{*}{Parameter} & \multirow{2}{*}{Unit} & Narrow CIE \\
 &  & $+$ Broad CIE model\\
\noalign{\smallskip}\hline\noalign{\smallskip}
\multicolumn{3}{c}{Component Em1}\\\noalign{\smallskip}
{\tt cie} $kT$ $^{(1)}$ & keV & $8\pm1$ \\\noalign{\smallskip}
{\tt cie} $n_{\rm H} n_{\rm e}V$ $^{(2)}$ & $10^{63}$ cm$^{-3}$ & $1.7^{+1.0}_{-0.5}$ \\\noalign{\smallskip}
{\tt reds} $v_{\rm shift}$ $^{(3)}$ & km s$^{-1}$ & $2600\pm100$ \\\noalign{\smallskip}
{\tt vgau} $v_{\rm broad}$ $^{(4)}$ & km s$^{-1}$ & $400^{+200}_{-100}$ \\\noalign{\smallskip}
\hline\noalign{\smallskip}
\multicolumn{3}{c}{Component Em2}\\\noalign{\smallskip}
{\tt cie} $kT$ $^{(1)}$ & keV & $7.5^{+0.7}_{-0.6}$ \\\noalign{\smallskip}
{\tt cie} $n_{\rm H} n_{\rm e}V$ $^{(2)}$ & $10^{63}$ cm$^{-3}$ & $3.0^{+0.4}_{-0.5}$ \\\noalign{\smallskip}
{\tt reds} $v_{\rm shift}$ $^{(3)}$ & km s$^{-1}$ & $-1500\pm100$ \\\noalign{\smallskip}
{\tt vgau} $v_{\rm broad}$ $^{(4)}$ & km s$^{-1}$ & $590^{+110}_{-90}$ \\\noalign{\smallskip}
\hline\noalign{\smallskip}
\multicolumn{3}{c}{Component Em3}\\\noalign{\smallskip}
{\tt cie} $kT$ $^{(1)}$ & keV & $7.3^{+0.8}_{-0.9}$ \\\noalign{\smallskip}
{\tt cie} $n_{\rm H} n_{\rm e}V$ $^{(2)}$ & $10^{63}$ cm$^{-3}$ & $3.3^{+0.9}_{-1.3}$ \\\noalign{\smallskip}
{\tt reds} $v_{\rm shift}$ $^{(3)}$ & km s$^{-1}$ & $4800^{+1400}_{-800}$ \\\noalign{\smallskip}
{\tt vgau} $v_{\rm broad}$ $^{(4)}$ & km s$^{-1}$ & $2500^{+800}_{-900}$ \\\noalign{\smallskip}
\hline\noalign{\smallskip}
$C$-stat / d.o.f. &  & 3653.1 / 3346 \\\noalign{\smallskip}
AICs &  & 3701.5 \\\noalign{\smallskip}
\hline
\end{tabular}
\begin{tablenotes}
\item
(1) The electron and ion temperatures. 
(2) Normalization of the {\tt cie} model. 
(3) Doppler shift velocity of the {\tt cie} model. Positive values represent inflow, and negative values represent outflow.
(4) Broadening velocity included the effect of the thermal broadening. 
\end{tablenotes}
\end{threeparttable}
\end{table}

\subsubsection{Combination of the photo-ionized emission and collisional ionized emission}
Finally, we tried a combination of the {\tt pion} and {\tt cie} models. 
There are two possible combinations. 
First, we tried the broad {\tt cie} $+$ narrow {\tt pion} $\times2$ model. 
However, the {\tt cie} component could not explain the broad component for the same reason described in the Sect. \ref{cieFit}. 
The narrow {\tt pion} model components attempted to fit the residuals of the broad component instead of the narrow features.
Therefore, the broad {\tt cie} $+$ narrow {\tt pion} $\times2$ model was deemed unsuitable. Next, we tried the broad {\tt pion} $+$ narrow {\tt cie} $\times2$ model. 
This model successfully explained the $6.5-7.0$ keV excess. 
The model for fitting the ionized Fe-K emission lines is
\begin{eqnarray}\label{A1model}
{\tt hot\times reds\times \{hot \times pion_{abs} \times pow}\nonumber\\
{\tt  + \sum_{i=1}^{2}gaus_i + \sum_{i=1}^{3}(MyTorus_i \times pow \times spei_i) }\nonumber\\
{\tt + \sum_{i=1}^{2}(cie\,_i\times reds\,_i \times vgau_i)\} }\nonumber\\
{\tt + pion\times pow \times vgau \} }
\end{eqnarray}
We refer to this as the Narrow CIE $+$ Broad PION model.

The fitting results of the Narrow CIE $+$ Broad PION model are shown in Fig. \ref{CIE}, and the best-fit parameters are summarized in Table \ref{CIEPIONTable}. 
The C-statistics and AICs prefer the Narrow CIE $+$ Broad PION model, yielding lower values than those of the Narrow PION $+$ Broad PION model. 
For the broad component, the parameters are consistent between the Narrow CIE $+$ Broad PION model and the Narrow PION $+$ Broad PION model.
The redshift and broadening velocity of the narrow two {\tt cie} components are also consistent with those of the Narrow PION $+$ Broad PION model. 
Both the {\tt pion} and {\tt cie} models could explain the narrow ionized component (especially for Em2); however, they produce different model curves. 
From Fig. \ref{CIE}, the differences for the Em2 component are that the {\tt cie} model predicts a stronger Fe {\sc xxv} w line (which fits the data well), while the {\tt pion} model predicts a stronger Fe {\sc xxvi} line.
In a photo-ionization scenario, the Fe {\sc xxv} w line is typically expected to be weaker than in collisional ionization.
However, assuming the different line-of-sight for the direct power-law emission and ionized component, the Fe {\sc xxvi} w line produced by the photo-ionization can be as strong as that from collisional ionization \citep[e.g.][]{Wojdowski2003, Bianchi2005, Porquet2010}. 
This is because with the different line-of-sight, the ionized component does not absorb the power-law emission, and we would instead observe scattered continuum and recombination emission. 
In addition, in photo-ionization, the w line intensity decreases with increasing column density \citep{Bianchi2005, Chakraborty2021}. 
As a result, we cannot statistically distinguish between the pure PION model and the Narrow CIE $+$ Broad PION model; thus, we will discuss both cases below.

\begin{table}[htbp]
 \caption{Best-fit parameters of the Narrow CIE $+$ Broad PION model. }
 \label{CIEPIONTable}
 \small\centering
  \begin{threeparttable}
  \begin{tabular}{lcc}
\noalign{\smallskip}\hline\hline\noalign{\smallskip}
\multirow{2}{*}{Parameter} & \multirow{2}{*}{Unit} & Narrow CIE $+$ \\
 &  & Broad PION model\\
\noalign{\smallskip}\hline\noalign{\smallskip}
\multicolumn{3}{c}{Component Em1}\\\noalign{\smallskip}
{\tt cie} $kT$ $^{(1)}$ & keV & $5^{+1}_{-2}$ \\\noalign{\smallskip}
{\tt cie} $n_{\rm H} n_{\rm e}V$ $^{(2)}$ & $10^{63}$ cm$^{-3}$ & $1.3^{+0.4}_{-0.3}$ \\\noalign{\smallskip}
{\tt reds} $v_{\rm shift}$ $^{(3)}$ & km s$^{-1}$ & $2600\pm100$ \\\noalign{\smallskip}
{\tt vgau} $v_{\rm broad}$ $^{(4)}$ & km s$^{-1}$ & $400\pm100$ \\\noalign{\smallskip}
\hline\noalign{\smallskip}
\multicolumn{3}{c}{Component Em2}\\\noalign{\smallskip}
{\tt cie} $kT$ $^{(1)}$ & keV & $7.5^{+0.8}_{-0.7}$ \\\noalign{\smallskip}
{\tt cie} $n_{\rm H} n_{\rm e}V$ $^{(2)}$ & $10^{63}$ cm$^{-3}$ & $2.8^{+0.5}_{-0.4}$ \\\noalign{\smallskip}
{\tt reds} $v_{\rm shift}$ $^{(3)}$ & km s$^{-1}$ & $-1500\pm100$ \\\noalign{\smallskip}
{\tt vgau} $v_{\rm broad}$ $^{(4)}$ & km s$^{-1}$ & $550^{+110}_{-80}$ \\\noalign{\smallskip}
\hline\noalign{\smallskip}
\multicolumn{3}{c}{Component Em3}\\\noalign{\smallskip}
{\tt pion} $N_{\rm H}$ $^{(5)}$ & $10^{23}$ cm$^{-2}$ & $1.8^{+0.5}_{-0.4}$ \\\noalign{\smallskip}
{\tt pion} $\log \xi$ $^{(6)}$ & erg cm s$^{-1}$ & $3.23^{+0.05}_{-0.04}$ \\\noalign{\smallskip}
{\tt pion} $v_{\rm shift}$ $^{(7)}$ & km s$^{-1}$ & $4000\pm500$ \\\noalign{\smallskip}
{\tt vgau} $v_{\rm broad}$ $^{(8)}$ & km s$^{-1}$ & $3100\pm500$ \\\noalign{\smallskip}
\hline\noalign{\smallskip}
$C$-stat / d.o.f. &  & 3617.6 / 3346 \\\noalign{\smallskip}
AICs &  & 3666.0 \\\noalign{\smallskip}
\hline
\end{tabular}
\begin{tablenotes}
\item
(1) The electron and ion temperatures. 
(2) Normalization of the {\tt cie} model. 
(3) Doppler shift velocity of the {\tt cie} model. Positive values represent inflow, and negative values represent outflow.
(4) Broadening velocity. 
(5) Column density of the {\tt pion} component. 
(6) Logarithm of the ionization parameter of the {\tt pion} component. 
(7) Flowing velocity of the {\tt pion} component, a positive value means the inflow and a negative value means the outflow. 
(8) Broadening velocity of the {\tt vgau} component which included the thermal broadening. 
The turbulence velocity of the {\tt pion} component was fixed to $100$ km s$^{-1}$, and the covering factor for emission was fixed to $0.2$. 
\end{tablenotes}
\end{threeparttable}
\end{table}

\subsection{Modeling of the absorption line features}
\subsubsection{PION modeling}
\begin{figure}[htbp]
 \centering
 \includegraphics[width=\hsize]{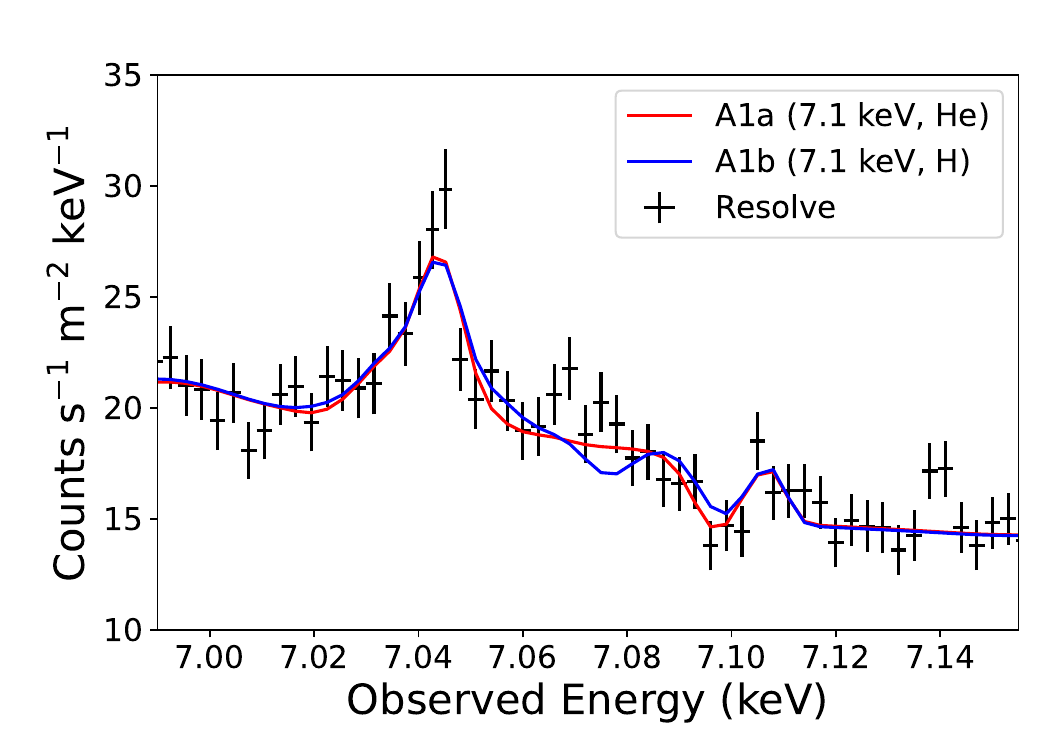}
 \caption{Resolve spectra ($6.9-7.1$ keV) and model curve of the A1a model and A1b model, wherein the black point, the red line, and the blue line represent the Resolve spectra, A1a model, and A1b model, respectively. }
 \label{A1_7keV}
\end{figure}

\begin{figure}[htbp]
 \centering
 \includegraphics[width=\hsize]{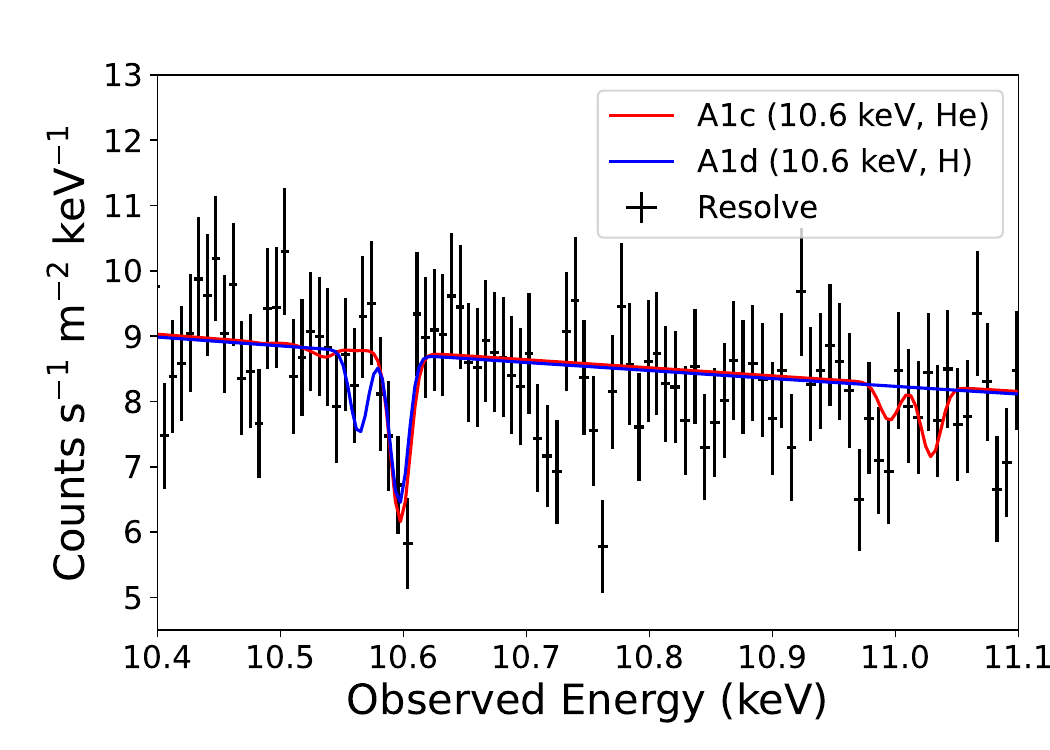}
 \caption{Resolve spectra ($10.4-11.1$ keV), model curve of the A1d model and A1c model, wherein the black point, the red line, and the blue line represent the Resolve spectra and A1d model and A1c, respectively. The Resolve spectrum is binned by a factor of 2 after the optimal binning for visual clarity.}
 \label{A1_10keV}
\end{figure}

Table \ref{line4sig} shows two absorption lines above $7$ keV (at 7.1 keV and 10.6 keV). 
In this subsection, we modeled them with the {\tt pion} absorption model \citep{Miller2015, Mehdipour2016}. 
Spectral fitting was performed by adding a {\tt pion} absorption model to the Narrow CIE $+$ Broad PION model described above to save computing time.
For the {\tt pion} absorption modeling, the free parameters were the hydrogen column density $N_{\rm H}$ (cm$^{-2}$), ionization parameter $\log \xi$ (erg cm s$^{-1}$), outflow velocity $v_{\rm out}$ (km s$^{-1}$) and turbulence velocity $v_{\rm turb}$ (km s$^{-1}$). 
We started with a single-zone photoionized absorber to express these absorption lines individually (Table \ref{absPION}). 
The model is 
\begin{eqnarray}\label{A1model}
{\tt hot\times reds\times \{hot \times pion_{abs} \times pow\nonumber}\\
{\tt + \sum_{i=1}^{2}gaus_i + \sum_{i=1}^{3}(MyTorus_i \times pow \times spei_i) }\nonumber\\
{\tt + \sum_{i=1}^{2}(cie\,_i\times reds\,_i \times vgau_i) }\nonumber\\
{\tt + pion\times pow \times vgau \} }
\end{eqnarray}
When leaving the covering factor $C_{\rm f}$ of the absorber free, the $C$-stat did not improve. Therefore, $C_{\rm f}$ of all absorption {\tt pion} was fixed to 1. 
Table \ref{absPION} summarizes the result of fitting only the 7.1 keV line (A1a model, Fig. \ref{A1_7keV}) and only the 10.6 keV line (Fig. \ref{A1_10keV}). 
There are two fitting results for each 7.1 keV and 10.6 keV absorption line. 
One is the H-like Fe (Fe {\sc xxvi}) case (A1b model and A1d model), and another is the He-like Fe (Fe {\sc xxv}) case (A1a model and A1c model). 
The above four models improved the {\it C}-statistics and the AIC significantly over the Narrow CIE $+$ Broad PION model, and thus it can be considered that these {\tt pion} components are significant. 
We found the wind outflows in a jet-dominated object. 
In the A1b and A1d models (H-like Fe case), the column density of the second {\tt MyTorus} moved to the lower limit ($1.1\times10^{22}$ cm$^{-2}$). 
This is because the column density of {\tt pion} is as large as $2\times10^{24}$ cm$^{-2}$ and thus the incident power-law component should be heavily absorbed.
Then, the best-fit power-law normalization became large by a factor of $\sim$10.
On the other hand, the line flux of the {\tt MyTorus} is not affected by the {\tt pion} absorber.
Consequently, the relative line flux ratio to the incident power-law becomes smaller.
This normalization is also tied to that of {\tt MyTorus}, and thus the column density is forced to become smaller to reproduce the small line flux ratio.

When compared statistics, the A1b model was better than the A1a model ($\Delta C=-72.5$ and $\Delta {\rm AIC} =-72.5$), and the A1d model was better than the A1c model ($\Delta C=-51.4$ and $\Delta {\rm AIC} =-51.4$). 
When we compared with the $C$-stat of the Resolve, the A1b model was better than the A1a model ($\Delta C=-9.6$), and the A1d model was better than the A1c model ($\Delta C=-1.2$). 
Then, we compared the $C$-stat of the limited band for Resolve. 
For the A1a and A1b model, the $C$-stat of the $6.5-7.6$ band keV was 436.7 and 441.6. 
When we focused on the 7.1 keV absorption line, the A1a (He-like) model was preferred; however, the $C$-stat of the total Resolve band preferred the A1b (H-like) model. 
This is because the $C$-stat of the $5-6.2$ keV band was 518.5 (A1a model) and 487.1 (A1b model), and the A1b model is significantly preferred ($\Delta C=-31.3$). 
A similar trend is present even for the 10.6 keV absorption line.
The $C$-stat of the $9.5-12$ keV band for the A1c and A1d model was 632.8 and 635.0. 
However, for the A1c and A1d model, the $C$-stat of the $5-6.2$ band keV was 520.3 and 494.6. 
As a result, when we focused on the absorption lines, the He-like case (A1a and A1c model) is preferred, but the $5-6.2$ keV band preferred the H-like case (A1b and A1d model).

$C$-stat of the Resolve for each model is 3179.4 (A1a), 3169.8 (A1b), 3181.4 (A1c), and 3180.1 (A1d). 
The total $C$-stat of the H-like case significantly improved than the He-like case (for both 7.1 keV and 10.6 keV absorption lines); however, the $C$-stat of the Resolve is less different.
This suggests there are significant differences in the NuSTAR spectra. 
There is no significant difference in the model curve within the Resolve band (Fig. \ref{A1cdDiffModel}). 
The slight differences lie in the presence of weaker absorption lines in the Resolve band and above the 40 keV band. 
This is because the A1d model has a very large column density, so even at high ionization, the continuum is reduced by scattering.
However, increasing the power-law normalization by a factor of 10 to counteract the scattering effect renders the two model curves visually indistinguishable.
However, as we discuss in subsection \ref{Wind}, the upper limit of position for the A1b model and A1d models (Fe {\sc xxvi} case) is inside the ISCO.

As a result, we could not simply determine whether Fe {\sc xxv} and Fe {\sc xxvi} for these absorption lines. 
The total $C$-stat preferred Fe {\sc xxvi}; however, the $C$-stat of the limited band for Resolve, which is focused on the absorption line, and their radius preferred Fe {\sc xxv}.

Next, we tried the combination of two components.  
When the multiple components of the {\tt pion} model are considered, the first component sees the non-absorbed SED, and the second component sees the absorbed SED (absorbed by the first component). 
The position setting for the multiple components of the {\tt pion} model component was determined with the faster 10.6 keV component positioned inside. 
We tested four cases, Fe {\sc xxv} and Fe {\sc xxvi}, for each of the two absorption lines.
The model for the combination of two {\tt pion} absorption components is expressed as 
\begin{eqnarray}\label{A2model}
{\tt hot\times reds\times \{hot \times pion_{abs} \times pion_{abs} \times pow}\nonumber\\
{\tt  + \sum_{i=1}^{2}gaus_i + \sum_{i=1}^{3}(MyTorus_i \times pow \times spei_i) }\nonumber\\
{\tt + \sum_{i=1}^{2}(cie\,_i\times reds\,_i \times vgau_i) }\nonumber\\
{\tt + pion\times pow \times vgau \} }
\end{eqnarray}
This model is simply the combination of the two absorbers. 
The fitting results about the combination of two {\tt pion} absorption components are shown in Table \ref{absPION}. 
The A2a model means the combination of the A1a and A1c model, the A2b model means the combination of the A1a and A1d model, the A2c model means the combination of the A1c and A1c model, and the A2d model means the combination of the A1b and A1d model. 
From the $C$-stat and AICs, the combination of the two {\tt pion} absorption components gave a significantly better fit than the one {\tt pion} absorption component. 


\begin{table*}[htbp]
 \caption{Best-fit of {\tt pion} absorption models. }
 \label{absPION}
 \centering
 \begin{threeparttable}\small
  \begin{tabular}{lcccccccc}
\noalign{\smallskip}\hline\hline\noalign{\smallskip}
 Model & Comp. & $N_{\rm H}$ & $\log \xi$ & $v_{\rm out}$ & $v_{\rm turb}$  & C-stat & d.o.f. & AIC\\\noalign{\smallskip}
  &  & ($10^{21}$ cm$^{-2}$) & (erg cm s$^{-1}$) & (km s$^{-1}$) & (km s$^{-1}$) \\
\noalign{\smallskip}\hline\noalign{\smallskip}
A1a & 7.1\,keV (He) & $1.5\pm0.4$ & $2.91^{+0.09}_{-0.07}$ & $-17790^{+80}_{-60}$ & $160^{+90}_{-70}$ & 3596.0 & 3342 & 3652.5 \\\noalign{\smallskip}
A1b & 7.1\,keV (H) & $(2.1\pm0.1)\times10^{3}$ & $5.9\pm0.1$ & $-5890\pm70$ & $<230$ & 3523.5 & 3342 & 3580.0 \\\noalign{\smallskip}
A1c & 10.6\,keV (He) & $5\pm2$ & $3.3\pm0.1$ & $-129010^{+50}_{-60}$ & $200^{+70}_{-50}$ & 3591.2 & 3342 & 3647.7 \\\noalign{\smallskip}
A1d & 10.6\,keV (H) & $(2.0\pm0.1)\times10^{3}$ & $5.8\pm0.1$ & $-119080^{+60}_{-70}$ & $160\pm70$ & 3539.8 & 3342 & 3596.3 \\
\noalign{\smallskip}\hline\noalign{\smallskip}
\multirow{2}{*}{A2a}  & 7.1\,keV (He) & $1.5\pm0.4$ & $2.91^{+0.09}_{-0.07}$ & $-17780^{+80}_{-60}$ & $160^{+90}_{-70}$ & \multirow{2}{*}{3569.3} & \multirow{2}{*}{3338} & \multirow{2}{*}{3633.3}\\\noalign{\smallskip}
 & 10.6 keV (He) & $5\pm2$ & $3.3\pm0.1$ & $-129010^{+50}_{-60}$ & $200^{+70}_{-50}$ \\\noalign{\smallskip}
\multirow{2}{*}{A2b}  & 7.1\,keV (He) & $1.6^{+0.5}_{-0.4}$ & $2.89\pm0.08$ & $-17790^{+70}_{-60}$ & $160^{+80}_{-70}$ &  \multirow{2}{*}{3516.9} & \multirow{2}{*}{3338} & \multirow{2}{*}{3581.5}\\\noalign{\smallskip}
& 10.6 keV (H) & $(2.1\pm0.1)\times10^3$ & $5.8\pm0.1$ & $-119010^{+60}_{-70}$ & $160\pm70$ & \\\noalign{\smallskip}
\multirow{2}{*}{A2c} & 7.1\,keV (H) & $5^{+4}_{-2}$ & $3.4^{+0.2}_{-0.1}$ & $-5860\pm50$ & $<190$ & \multirow{2}{*}{3604.8} & \multirow{2}{*}{3338} & \multirow{2}{*}{3669.4} \\\noalign{\smallskip}
 & 10.6 keV (He) & $(2.0^{+0.1}_{-0.2})\times10^3$ & $5.8\pm0.1$ & $-119080^{+60}_{-70}$ & $160\pm70$\\\noalign{\smallskip}
\multirow{2}{*}{A2d} & 7.1\,keV (H) & $(2.1\pm0.1)\times10^3$ & $5.9\pm0.1$ & $-5890\pm70$ & $<230$ & \multirow{2}{*}{3505.5} & \multirow{2}{*}{3338} & \multirow{2}{*}{3561.9} \\\noalign{\smallskip}
 & 10.6 keV (H) & $(0.2^{+1.1}_{-0.1})\times10^2$ & $4.1^{+0.4}_{-0.2}$ & $-119060\pm70$ & $190^{+70}_{-60}$\\\noalign{\smallskip}
\hline\noalign{\smallskip}
\end{tabular}
\begin{tablenotes}
\item 
$N_{\rm H}$: Column density.  $\log \xi$: Logarithm of the ionization parameter. $v_{\rm out}$: Outflow velocity. $v_{\rm turb}$: Turbulence velocity. 
\end{tablenotes}
\end{threeparttable}
\end{table*}

To evaluate whether NuSTAR data drives the preference for certain models. 
We removed the NuSTAR spectrum from the analysis and fitted the spectrum with Resolve only ($3-15$ keV). 
Then, the parameters of the ionized emission changed (specifically, the broadening velocity of the narrow (redshift) component increased from 400 km s$^{-1}$ to 2000 km s$^{-1}$). 
This is because, when using only the Resolve data ($3-15$ keV), since hard X-rays are excluded, the continuum shape is altered to account for the excess in the Fe-K band.
Then, we fixed the broadening velocities of the ionized emission components and fit the Resolve spectrum using the A1a and A1c models. 
When the A1a model is applied, the photon index $\Gamma = 1.858$ changes to $\Gamma = 1.915$, and when the A1c model is applied, the photon index $\Gamma = 1.857$ changes to $\Gamma = 1.909$.
However, the parameters of the absorption components are not changed significantly. 
As a result, the broadening velocity of the narrow component is sensitive to the continuum shape. 
Thus, some uncertainties in the distance estimation of the narrow ionized emission component.

\subsubsection{Fe-K absorption edge modeling for 7.1 keV absorption line feature}
\begin{figure}[htb]
 \centering
 \includegraphics[width=\hsize]{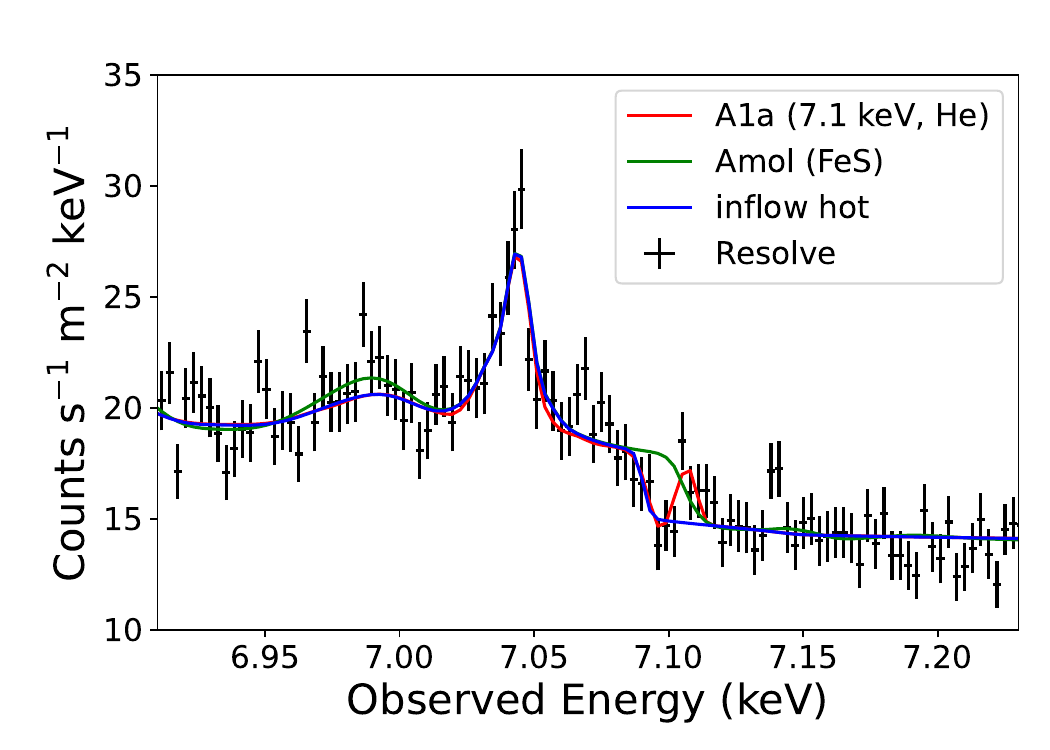}
 \caption{XRISM/Resolve ($6.95-7.25$ keV) spectra with the model curve of the A1a model, {\tt amol} model, and inflowing {\tt hot} model, wherein the black point, the red line, the green line, and the blue line represent the Resolve spectra, the model curve of the A1a model, {\tt amol} model, and inflowing {\tt hot} model, respectively.}
 \label{hot_amol}
\end{figure}

The 7.1 keV absorption line can be superimposed on the Fe-K edge structure. Therefore, it could appear as an artifact if the Fe-K edge is not well modeled; that is, if the model shape does not accurately reproduce the edge energy and shape.
Here, we investigated two alternative cases for Fe-K edge modeling: the {\tt amol} model for the dust (molecular) absorption and a Doppler-shifted {\tt hot} model. 
The {\tt amol} model calculates transmission through molecular absorbers \citep{Pinto2010, Rogantini2018}. 
We applied the {\tt amol} model to the power-law component. 
For the {\tt amol} fitting, the range of Fe abundance in the {\tt hot} model was set to $0-0.2$, the assumption that up to 20\% of Fe is not contained in the dust. This is a typical value for the Milky Way \citep[e.g.][]{Psaradaki2023}. 
We can choose molecules in the {\tt amol} model that could produce a Fe-K edge, and we tried all of them. 
As a result, the choice of Troilite (FeS) molecule gave the best $C$-stat and AICs ($C$-stat is 3601.0 and AICs is 3653.4). 
The best-fit parameters are summarized in Table \ref{edge}.
However, compared to the A1a model, the statistics were not favorable, and thus the {\tt amol} modeling could not account for the 7.1 keV absorption line (-like) structure (Fig. \ref{hot_amol}). 
As seen in the figure, the {\tt amol} model gives a lower edge energy than {\tt hot} and a slightly different edge shape, but cannot explain the 7.1 keV absorption line feature. 
There are various dust models \citep[e.g.,][]{Ricci2023}, but the trend is almost similar; the edge energy decreases with a slight change in edge shape. 
Therefore, we consider that these models do not explain the 7.1 keV absorption line feature.

Next, we tried the Doppler-shifted {\tt hot} model to change the edge energy by allowing the shift velocity of {\tt hot} in the Narrow CIE $+$ Broad PION model to vary. 
As summarized in Table \ref{edge}, the redshift velocity becomes $v_{\rm flow}=+810^{+170}_{-90}$ km s$^{-1}$. 
The $C$-stat was 3603.2, and the AIC was 3635.6. 
When we compared this model with the  A1a {\tt pion}  model, the $\Delta$AIC was 1.1 (slightly favoring the A1a model). 
The column density of this model was $(1.655^{+0.008}_{-0.007})\times10^{23}$ cm$^{-2}$; therefore, we considered that this absorption originates from the torus \citep[e.g.,][]{Fukazawa2011}.
This indicates a fast inflow of torus material. 
On the other hand, there is no evidence of such inflow at other wavelengths for Cen A. 
\citet{Zhou2019} reported the detection of inflowing gas with 5,000 km s$^{-1}$ for quasars, suggesting that it is flowing at $\sim$1000 gravitational radius. 
The Eddington ratio of Cen A is as small as $\sim10^{-3}$, and such a massive inflow with a column density of $\sim10^{23}$ cm$^{-2}$ at a small radius is unlikely.  
Fig. \ref{hot_amol} shows that the inflowing {\tt hot} model cannot explain the excess count around 7.11 keV, and thus the {\tt pion} A1a model gives better statistics as seen in Table \ref{edge}.
Therefore, we considered the {\tt pion} model to be more suitable for the 7.1 keV structure. 

\begin{table}[htbp]
 \caption{Best-fit parameters of modeling the 7.1 keV structure. }
 \label{edge}
 \centering  
 \adjustbox{max width=\hsize,center}{
 \begin{threeparttable}
  \begin{tabular}{lccc}
\noalign{\smallskip}\hline\hline\noalign{\smallskip}
 & {\tt pion} (A1a) & {\tt amol} (FeS) & Inflow {\tt hot}\\
\noalign{\smallskip}\hline\noalign{\smallskip}
$N_{\rm H}$ ($\times10^{22}$ cm$^{-2}$) & $0.15\pm0.04$ & $(4.62\pm0.02)\times10^{-4}$ & $16.55^{+0.08}_{-0.07}$ \\ \noalign{\smallskip}
$v_{\rm flow}$ (km s$^{-1}$) & $-17790^{+80}_{-60}$ & $0$ & $+810^{+170}_{-90}$ \\
\noalign{\smallskip}\hline\noalign{\smallskip}
$C$-stat & 3596.0 & 3601.0 & 3603.2 \\ \noalign{\smallskip}
AICs & 3652.5 & 3653.4 & 3653.6 \\ 
\noalign{\smallskip}\hline\noalign{\smallskip}
\end{tabular}
\begin{tablenotes}
\item
\end{tablenotes}
 \end{threeparttable}}
\end{table}

\section{Discussion} \label{discuss}
XRISM/Resolve observations revealed that several ionized Fe-K emission or absorption lines with a shift velocity and a width in the range of $10^{2-5}$ km s$^{-1}$ are present in the X-ray spectrum of Cen A. 
Some of the ionized Fe-K emission lines are redshifted. 
To discuss their origins, we estimate physical quantities calculated from the obtained ionization parameter and column density for the {\tt pion} model or the emission measure for the {\tt cie} model. 

The ionization parameter is defined as in Eq. (\ref{eqr1}). 
The ionization luminosity ($L_{\rm ion}$) is calculated from the best-fit power-law model to be $5\times10^{42}$ erg s$^{-1}$.
Since the distance of the gas from the ionizing source ($r$) is larger than the depth $d$ of the gas along the line of sight ($r>d$) and the column density is $N_{\rm H}=nd$, we can obtain an upper limit on the distance as $r<\frac{L}{\xi N_{\rm H}}$. 
The mass flow rate is calculated as $\dot{M}_{\rm flow} = 4\pi\mu m_p r N_{\rm H} C_{\rm v} v$ \citep{Crenshaw2003,Crenshaw2012,Tombesi2017, Gallo2023}, where $\mu= 1.23$ is the mean atomic mass per proton, $m_p$ is the proton mass, and $v$ is the flow (redshift or blueshift) velocity. 
The covering factor $\Omega/4\pi$ is assumed to be 0.2 for the emission line. 
The kinetic power is $\dot{E}_{\rm kinetic}={1\over2}\dot{M}v^2$ \citep{Tombesi2013}.

In the case of the {\tt cie} model, assuming a typical size of the plasma as $R$, the number density of the gas is calculated as $n=\sqrt{\frac{EM}{qR^3}}$, where $EM$ is the emission measure and $q$ is a geometrical factor. 
Then, the total mass of the gas is $M_{\rm gas}=\mu m_p nqR^3$, the mass flow rate is $\dot{M}_{\rm flow} = \mu m_p nqR^3 v$, the kinetic power is $\dot{E}_{\rm kinetic}={1\over2}\dot{M}_{\rm flow}v^2$, and the thermal energy is $E_{\rm thermal}=nkTqR^3$.

Table \ref{massloss} summarizes the obtained parameters and the calculated plasma quantities, where we show the dependence on $r$, $C_{\rm v}$, or $q$, $R$. 
For the narrow emission lines, there are two components, but here we treat them in the same manner by using the parameters as shown in the table.
Note that the black hole mass of Cen A is $(5.5 \pm 3.0) \times 10^7 M_\odot$, and thus the Eddington luminosity is $6.9\times10^{45}$ erg s$^{-1}$. 
The $2-10$ keV luminosity of the power-law component is $2.07\times10^{42}$ erg s$^{-1}$ (derived from the A1a model after correcting for absorption). 
To estimate the bolometric luminosity, we use the $1-1000$ Ryd luminosity ($L_{\rm ion}$) of the power-law component as the lower limit and the bolometric correction factor $\kappa_{2-10}=20$ \citep{Vasudevan2007} relative to the $2-10$ keV power-law luminosity as the upper limit, because Cen A is a low Eddington ratio AGN. 
The $1-1000$ Ryd (13.6 eV $-$ 13.6 keV) luminosity ($L_{\rm ion}$) of the power-law component from the A1a model is $L_{\rm ion}=6.65\times10^{42}$ erg s$^{-1}$ (we could obtain the same luminosity using the A1c model). 
Thus, the upper limit of the bolometric luminosity is $L_{\rm bol}=4.1\times10^{43}$ erg s$^{-1}$, and the lower limit  is $L_{\rm bol}=6.7\times10^{42}$ erg s$^{-1}$. 
The Eddington ratio is estimated to be $\lambda=0.001-0.006$. 
The Eddington mass flow rate is $\dot{M}_{\rm Edd}=1.22M_\odot$ yr$^{-1}$.

\begin{table*}[htbp]
 \caption{Calculated physical quantities. }
 \label{massloss}
 \centering\hspace{-2cm}
  \adjustbox{max width=\textwidth,center}{
 \begin{threeparttable}
  \begin{tabular}{lcccccc}
\noalign{\smallskip}\hline\hline\noalign{\smallskip}
 & Unit & Broad component & \multicolumn{2}{c}{Narrow component} & UFO$_{7.1\,{\rm keV}}$ & UFO$_{10.6\,{\rm keV}}$ \\
\noalign{\smallskip}\hline\noalign{\smallskip}
Shift velocity & km s$^{-1}$ & 4600 & \multicolumn{2}{c}{2600, $-$1500} & $-$17790 & $-$129010 \\\noalign{\smallskip}
Broadening velocity & km s$^{-1}$ & 3300 & \multicolumn{2}{c}{400, 600} & 150 & 190 \\\noalign{\smallskip}
Column density & cm$^{-2}$ & $1.7\times10^{23}$ & $3\times10^{22}$ & - & $1.5\times10^{21}$ & $4\times10^{21}$ \\\noalign{\smallskip}
Ionization parameter & erg cm s$^{-1}$ & 3.2 & 3.1 & - & 2.9 & 3.3 \\\noalign{\smallskip}
Ionization mechanism &  & PIE & PIE & CIE &  \\\noalign{\smallskip}
Radius $^a$ & $R_{\rm S}$ & - & - & - & $500\left(\frac{n_{\rm H}}{10^{8} {\rm cm}^{-3}}\right)^{-0.5}$ & $330\left(\frac{n_{\rm H}}{10^{8} {\rm cm}^{-3}}\right)^{-0.5}$ \\\noalign{\smallskip}
Lower limit Radius $^b$ & $R_{\rm S}$ & - & - & - & $2.8\times10^2$ & $5.4$ \\\noalign{\smallskip}
Upper limit Radius $^c$ & $R_{\rm S}$ & $1300$ & $8\times10^3$ &  & $2.7\times10^5$ & $3.5\times10^4$ \\\noalign{\smallskip}
Mass-flow rate & $M_\odot$ yr$^{-1}$ & $0.01\left(\frac{r}{100 R_{\rm S}}\right)$ & $0.008\left(\frac{r}{10^3 R_{\rm S}}\right)$ & 0.04 $ q\left(\frac{r}{0.5 {\rm pc}}\right)$ & $2\times10^{-4}\left(\frac{r}{100 R_{\rm S}}\right)$ $\left(\frac{C_{\rm v}}{0.1}\right)$ & $4\times10^{-4}\left(\frac{r}{10 R_{\rm S}}\right)$ $\left(\frac{C_{\rm v}}{0.1}\right)$ \\\noalign{\smallskip}
Kinetic power & erg s$^{-1}$ & $1\times10^{41}\left(\frac{r}{100 R_{\rm S}}\right)$ & $1\times10^{41}\left(\frac{r}{10^3 R_{\rm S}}\right)$ & $5.3\times10^{40} q^{1.5}\left(\frac{r}{0.5 {\rm pc}}\right)^{1.5}$ & $4\times10^{40}\left(\frac{r}{100 R_{\rm S}}\right)$ $\left(\frac{C_{\rm v}}{0.1}\right)$ & $4\times10^{42}\left(\frac{r}{10 R_{\rm S}}\right)$ $\left(\frac{C_{\rm v}}{0.1}\right)$\\\noalign{\smallskip}
\hline
\end{tabular}
\begin{tablenotes}
\item
a: Radius, assuming the Keplerian motion using the broadening velocity. 
b: Lower limit of the radius using the $r_{\rm min}={2GM_{\rm BH} / v_{\rm out}^2}$. 
c: Upper limit of the radius using the $r_{\rm max}={{L_{\rm ion} C_{\rm v} / \xi N_{\rm H} }}$, where $C_{\rm v}=1$ is the volume filling factor. 
\end{tablenotes}
\end{threeparttable}}
\end{table*}

\subsection{Ionized emission lines}
We detected three ionized emission components in Cen A. 
XRISM has observed some AGNs in the PV phase, including two low-luminosity AGNs: M81* and Cen A
The M81* XRISM observation detected two ionized Fe-K emission components with widths of 800 km s$^{-1}$ and 200 km s$^{-1}$, and redshifted velocities of 170 km s$^{-1}$ and 1500 km s$^{-1}$, respectively \citep{Miller2025}. 
However, the ionization mechanism for the two components in M81* was not determined (photo-ionization or collisional ionization). 
In Cen A, both photo-ionized and collisionally ionized plasma models can represent the two narrow ionized components, while the origin of the broad ionized component is likely photo-ionization (as described by the {\tt pion} model). 
Such ionized emission line components might be a common feature of low-luminosity AGNs, but their physical properties differ between Cen A and M81*.

\subsubsection{Broad redshifted ionized emission} 
From the Table \ref{massloss}, the upper limit on the distance from the ionizing source is as small as $1300\,R_{\rm S}$. 
This is because the ionizing luminosity ($5\times10^{42}$ erg s$^{-1}$) is relatively low compared to the other XRISM observed UFO AGNs \citep[e.g.,][]{Xrism2025, Mehdipour2025, Noda2025}.  
In this case, the gravitational redshift is larger than 110 km s$^{-1}$ and thus cannot be ignored relative to the apparent redshift velocity of $4600$ km s$^{-1}$ (and would be larger if the distance were smaller).

Inflow, outflow, or rotation with a disk-like or spherical shape is expected to show a double-peak emission line profile, centered on the source-rest-frame line energy.
Therefore, if the flow is disk-like or spherical, an alternative interpretation is that the observed shift is purely gravitational (4600 km s$^{-1}$), which would place the emitting material at a distance from the central black hole of $r\sim30R_s$.
The line width of $\sim3000$ km s$^{-1}$ might represent a range of inflow or outflow velocity that blends the double-peak profile into a broad single peak. 
The escape velocity at this radius is $5.5\times10^4$ km s$^{-1}$, indicating that the flow would require supporting energy to keep the gas at this radius against the gravitational force (or if it were an outflow, it would fail to escape). 
A past period of high activity of the central engine might have ejected an outburst of gas, and its kinetic energy might now be supporting the gas.

One possibility of making a single line profile is that the blueshifted outflow component on the near side is obscured by the torus, leaving only the redshifted component on the far side visible. But $r$ is so small that the emitter will be seen to be almost point-like from the torus, and thus both the near and far sides would likely be either obscured or unobscured together.

Another possibility is that the flow is not disk-like or spherical, but the geometry is such that only redshifted emission is observed; for example, an inflow at a large angle relative to the disk. As discussed in \citet{Bogensberger2025}, the viewing angle of Cen A is uncertain but may be around 20--70 degrees. In that case, most of the flow region would show a redshift. One scenario to explain such a flow is a failed outflow \citet{Li2025}. In this case, the distance from the central black hole could be as large as $1300\,R_{\rm S}$ and the mass flow rate becomes 0.1 $M_{\odot}$ yr$^{-1}$.
A high activity in the past could have produced such a massive outflow. In summary, the broad emission component is not straightforwardly understood.

\subsubsection{Narrow ionized emission} 
XRISM/Resolve revealed both blueshifted and redshifted narrow ionized emission components in the Fe-K band, and their shift velocities are similar ($\sim$2000 km s$^{-1}$). We cannot rule out the possibility of two different velocity components, but here we consider them to share the same origin for simplicity, indicating that the gas geometry is disk-like or spherical.

If the gas is a photoionized plasma, the upper limit on its distance from the central ionizing source is $8\times10^3~R_{\rm S}$. The gravitational redshift is $\sim$21~$(8\times10^3~R_{\rm S}/r)$ km s$^{-1}$ and the escape velocity is $\sim3400~(8\times10^3~R_s/r)^{0.5}$ km s$^{-1}$. 
Thus, the gravitational redshift could explain the small difference in velocity between the redshifted and blueshifted components if the distance is as large as $r\sim10^3\,R_{\rm S}$. 
The observed velocity is not much lower than the escape velocity, and thus, either inflow, outflow, or rotation can be considered. 
Then, the mass flow rate would be 0.008 $M_\odot$ yr$^{-1}$ ($0.006 \dot{M}_{\rm Edd}$), close to that estimated from the current Cen A luminosity. 
The kinetic power of the flow is estimated to be $\dot{E}_{\rm kinetic} = 1.1\times10^{40}$ erg s$^{-1}$. This could be driven by the current AGN activity.

The upper limit on the mass flow rate of these components is estimated to be $\dot{M}_{\rm blue} = 0.006\, M_\odot$ yr$^{-1}$. The upper limit on the kinetic power is estimated to be $\dot{E}_{\rm blue} = 1\times10^{41}$ erg s$^{-1}$.

In the case of a collisional ionized plasma, we cannot constrain the distance from the center, but its emission must originate from a region obscured by the torus; otherwise, their emission should appear in the soft X-ray band without absorption. 
Therefore, the plasma is likely located within the torus region of $<1$ pc. 
In addition, since the thermal broadening is at most 50 km s$^{-1}$ for the temperature of 7 keV, other broadening mechanisms, such as turbulence, must contribute with 400 km s$^{-1}$. 
Assuming a typical size of $r=0.5$ pc, we obtain a gas number density of $n_{\rm blue}\sim3.0\times10^3$ cm$^{-3} q^{-1.5}(r/0.5~{\rm pc})^{-1.5}$, a total gas mass $M = 7.4\times10^{-2} q^{1.5}(r/0.5~{\rm pc})^{1.5} M_\odot$, and a total thermal energy $E_{\rm thermal}=2.9\times10^{52} q^{1.5}(r/0.5~{\rm pc})^{1.5}$ erg. 
The escape velocity is around 940 km s$^{-1}$, and thus outflow or rotation are also possibilities, in addition to inflow. 
In the outflow or rotation case, the plasma would be produced if a part of the accreting matter (BLR, torus, or others) was shock-heated by a massive AGN outflow. 
The total thermal energy can be supplied by the current AGN power within $10^3$ yr. 
The total mass is smaller than that expected for the torus region, suggesting that only a part of the circumnuclear matter might be heated.
The mass flow rate is 0.04 $ q(r/0.5~{\rm pc}) M_{\odot}$ yr$^{-1}$ (0.03 $\dot{M}_{\rm Edd}$) and the kinetic power is $5.3\times10^{40} q^{1.5}(r/0.5~{\rm pc})^{1.5}$ erg s$^{-1}$. 
Therefore, the current AGN activity is sufficient to drive the outflow.

Another possibility is an outer region of radiatively inefficient accretion flow \citep[RIAF;][]{Quataert1999} inflow since Cen A is a low Eddington AGN \citep[$L/L_{\rm Edd}\sim10^{-3}$;][]{Fukazawa2016, Nakatani2023, Iwata2024, Bogensberger2024}. 
While the innermost regions of the RIAF are expected to be at high temperature with fully ionized gas, the outer region could remain cool enough to allow Fe {\sc xxv} and Fe {\sc xxvi} to exist. 
The two narrow components might correspond to the inflowing RIAF wind \citep{Shi2021, Shi2025}. 
In the case of the RIAF inflow, the mass flow rate is estimated as $\dot{M} = 0.4 M_\odot$ yr$^{-1}$, which is too high for RIAF.

\subsection{Ionized absorption lines}\label{Wind}
The 10.6 keV component has a significantly faster outflowing velocity than the 7.1 keV component. 
Therefore, the 10.6 keV component should be located within the inner of the 7.1 keV component, and we set the relation of the model components so in the fitting. 
Considering the inner {\tt pion} component has a higher (or the same) Hydrogen number density, the inner component should have a higher ionization degree than the outer component. 
However, the 10.6 keV component has a significantly lower ionization degree than the 7.1 keV component in the A2d model.  
Thus, the A2d model (Fe {\sc xxvi} for both absorption lines) might not suit the model. 
Moreover, the inner component should be more highly ionized than the outer component. 
Therefore, the A2c model (Fe {\sc xxvi} for the 7.1 keV component and Fe {\sc xxv} for the 10.6 keV component) might also not suit the model.

Assuming that the wind has a faster outflow velocity $v_{\rm out}$ than the escape velocity, the lower limit on the position of the ionized wind was evaluated as $D_{\rm min}=2GM_{\rm BH} / v_{\rm out}^2$, using the outflow velocity $v_{\rm out}$. 
The evaluated upper limit and lower limit distances from the SMBH for all two {\tt pion} absorption model cases are listed in Table \ref{absPIONRadius}. 
The upper limit distance in the A2b, A2c, and A2d models is the inner ISCO; therefore, it is physically unlikely for wind to flow from that radius. 
In addition, the upper and lower limits in the A2b, A2c, and A2d models are inconsistent, which suggests that the winds are failed winds \citep[e.g.,][]{Proga2004} or that the A2b, A2c, and A2d models are not suitable for physical interpretation. 
This inconsistency (upper limit distances from the SMBH are inside the ISCO) is due to their high column density (above $10^{24}$ cm$^{-2}$). 
In the {\tt pion} model, the continuum component counteracts the scattering effect, increasing the power-law normalization by a factor of 10. 
This behavior is also seen in using the {\tt xabs} model in SPEX code. 
From the past broadband Cen A observations \citep{Borkar2021}, the bolometric luminosity was computed to $L_{\rm bol}=9.45\times10^{42}$ erg s$^{-1}$. 
In the A1a model (or A1c model; both cases have similar luminosity), the $2-10$ keV luminosity is consistent with past CCD observations (e.g., obtained using the {\it Swift}/BAT). 
However, the power-law normalization in the models considering the H-like absorption line became larger by a factor of $\sim$10 than that of the He-like absorption line case. 
Despite the ionization degree being very large at $\log\xi=5.8$ erg cm s$^{-1}$, a significant difference appears in the continuum flux because the Hydrogen column density is large ($N_{\rm H}=2\times10^{24}$ cm$^{-2}$), leading to Compton scattering.
From the power-law normalization of the A1b models (or the A1d model; both cases have similar luminosity), the $2-10$ keV luminosity is $1.1\times10^{43}$ erg s$^{-1}$ after correcting for absorption. 
This significantly high luminosity, with the models considering the H-like absorption line, might be physically unlikely. 
Although this might be model-dependent. 
While the scattering is treated as energy-independent Thomson attenuation, effectively acting as a grey absorber, without Klein-Nishina effects or Compton energy redistribution, and all scattered photons are treated as lost.
Such an approximation can overestimate continuum suppression at high column densities and may artificially require a higher power-law normalization to fit the data. 
As a result, the inferred $\sim 10$ times normalization increase could reflect limitations of the radiative transfer treatment rather than a physical inconsistency of the Fe {\sc xxvi} solution. 
As a result, we do not reject the Fe {\sc xxvi}. 
Since the A2b, A2c, and A2d models using the {\tt pion} model might be physically unlikely. 
Hereafter, for the discussion, we use parameters from the model considering the He-like absorption line.

\begin{table*}[htbp]
 \caption{Evaluated radius from the absorption models. }
 \label{absPIONRadius}
 \centering\hspace{-2cm}
  \adjustbox{max width=\textwidth,center}{
 \begin{threeparttable}\small
  \begin{tabular}{lcccccccc}
\noalign{\smallskip}\hline\hline\noalign{\smallskip}
 & \multicolumn{2}{c}{A2a} & \multicolumn{2}{c}{A2b} & \multicolumn{2}{c}{A2c} & \multicolumn{2}{c}{A2d} \\\noalign{\smallskip}
 & 7.1 keV (He) & 10.6 keV (He) & 7.1 keV (He) & 10.6 keV (H) & 7.1 keV (H) & 10.6 keV (He) & 7.1 keV (H) & 10.6 keV (H) \\\noalign{\smallskip}
 \noalign{\smallskip}\hline\noalign{\smallskip}
$L_{\rm ion} / \xi$ $^{(1)}$ & 642 & 278 & 162 & 5.25 & 225 & 5.21 & 4.37 & 265 \\\noalign{\smallskip}
$N_{\rm H}$ $^{(2)}$ & $1.5\pm0.4$ & $5\pm2$ & $1.6^{+0.5}_{-0.4}$ & $(2.1\pm0.1)\times10^3$ & $5^{+4}_{-2}$ & $(2.1\pm0.1)\times10^3$ & $(2.1\pm0.1)\times10^3$ & $(0.2^{+1.1}_{-0.1})\times10^2$ \\\noalign{\smallskip}
$v_{\rm out}$ $^{(3)}$ & $-17780^{+80}_{-60}$ & $-129010^{+50}_{-60}$ & $-17790^{+70}_{-60}$ & $-119010^{+60}_{-70}$ & $-5860\pm50$ & $-119080^{+60}_{-70}$ & $-5890\pm70$ & $-119060\pm70$ \\\noalign{\smallskip}
$D_{\rm min}$ $^{(4)}$ & $(2.84^{+0.03}_{-0.04})\times10^2$ & $5.400^{+0.004}_{-0.005}$ & $(2.84\pm0.02)\times10^2$ & $6.346^{+0.006}_{-0.007}$ & $(2.62^{+0.05}_{-0.04})\times10^3$ & $6.338^{+0.006}_{-0.007}$ & $(2.59\pm0.06)\times10^3$ & $6.340\pm0.007$ \\\noalign{\smallskip}
$D_{\rm max}$ $^{(5)}$ & $(2.6^{+1.0}_{-0.6})\times10^5$ & $(3^{+2}_{-1})\times10^4$ & $(6^{+2}_{-1})\times10^4$ & $1.54^{+0.08}_{-0.07}$ & $(3^{+2}_{-1})\times10^4$ & $1.53^{+0.08}_{-0.07}$ & $1.28\pm0.06$ & $(8^{+8}_{-7})\times10^3$ \\\noalign{\smallskip}
\hline\noalign{\smallskip}
\end{tabular}
\begin{tablenotes}
\item 
(1) The ionized luminosity in unit of $10^{37}~{\rm cm^{-1}}$. This value is automatically calculated by the {\tt pion} model. The 10.6 keV component is looking at the unabsorbed SED, and the 7.1 keV component is looking at the absorbed SED (absorbed by the 10.6 keV component). 
(2) Column density in units of $10^{21}$ cm$^{-2}$. 
(3) Outflow velocity in units of km s$^{-1}$. 
(4) The lower limit on the position of the ionized wind. It was evaluated as $2GM_{\rm BH} / v_{\rm out}^2$ in units of $R_{\rm S}$. 
(5) The upper limit on the position of the ionized wind. It was evaluated as $(L/\xi) \times N_{\rm H}$ in units of $R_{\rm S}$. 
\end{tablenotes}
\end{threeparttable}}
\end{table*}

The Eddington ratio is estimated to be $\lambda=0.001-0.006$ (Sect. \ref{discuss}). 
This Eddington ratio is too small for the line-force mechanism \citep[e.g.,][]{Sim2010, Nomura2017}. 
On the other hand, the outflow velocity is too high for a thermally driven wind \citep[e.g.,][]{Faucher2012, Mizumoto2019}. 
Thus, either line-driven or thermal-driven wind launch mechanism is unlikely to be. 
For these reasons, the wind driving mechanism is thought to be Magneto-Hydrodynamics (MHD) driving. 
In this case, the shape of the absorption line is predicted to be asymmetric \citep{Gandhi2022, Tomaru2023}, but current data do not allow us to study this possibility.

From the simulations, if the ratio of wind kinetic power to their bolometric luminosity is above $\sim$5\% (and can be as low as $\sim$0.5\%), the wind can contribute to host galaxy feedback \citep[e.g.][]{Matteo2005}. 
In addition, quasi-analytical studies suggest that disk winds may promote the collimation of jets near SMBHs \citep{Fukumura2014, Globus2016, Blandford2019}.
The wind kinetic power to bolometric luminosity ratio are $\simeq 1.4C_{\rm v}\,\%$ for the 7.1 keV absorption line and $\simeq 27C_{\rm v}\,\%$ for the 10.6 keV absorption line. 
As discussed in the previous subsection, various possibilities of outflows are inferred from the emission lines; these absorbers might be a part of such outflow components, but with an extremely high velocity. 
In that case, a large covering factor ($>0.1$) leads to detectable emission lines. 
However, there is no evidence of corresponding emission lines, implying $C_{\rm v}<0.1$.
Thus, the wind from Cen A might not affect the feedback. 
On the other hand, the total power of the jet is estimated to be $P_{\rm jet} = (1-2)\times10^{43}$ erg s$^{-1}$, and an instantaneous jet power of $P_{\rm jet} = 6.6\times10^{42}$ erg s$^{-1}$ \citep[e.g.][]{Croston2009, Wykes2013, Neff2015}. 
The wind kinetic power is much lower than the jet power; therefore, the wind likely does not affect the jet collimation.

\citet{Yamada2024} reported a velocity gap at $\log v_{\rm out}$ [km s$^{-1}$]$\sim4$. 
They suggested that the gap might be an observational bias, as absorption lines with velocities of $\log v_{\rm out}$ [km s$^{-1}$]$\sim4$ would appear around the Fe-K edge and Fe-K$\beta$ emission line, and thus could not be identified with the CCD resolution. 
In this observation, the 7.1 keV absorption line appeared precisely in the Fe-K edge structure, but XRISM/Resolve resolved the Fe-K edge and the absorption line. 
This demonstrates the excellent capability of XRISM/Resolve for line searches.

On the other hand, the 10.6 keV component has a very fast velocity ($\log v_{\rm out}$ [km s$^{-1}$]$=5.11$). 
\citet{Yamada2024} compiled a list of AGNs that have UFOs with very fast outflow velocities ($\log v_{\rm out}$ [km s$^{-1}$]$\geq5$). 
The list of AGNs with very fast outflow velocity in \citet{Yamada2024} is in Table \ref{ufolist}. 
Most AGNs with outflow velocities above $\log v_{\rm out}$ [km s$^{-1}$]$\geq5$ were quasars.

\subsection{Relation with a large-scale outflow discovered by JWST}
JWST/MIRI observations of the Cen A core detected the outflowing gas via ionized Ar {\sc ii} and Ne {\sc iii} lines. They are narrow ($\sigma\sim600$ km s$^{-1}$) with velocity reaching approximately $+$1000 and $-$1400 km s$^{-1}$, respectively, within a $\sim10^{6}~R_{\rm S}$ (6 pc) region \citep{Alonso2025}. 
They calculated the upper limit of the kinetic power to be $1.3\times10^{42}$ erg s$^{-1}$.
If the distance of this outflow from the center is around 1 pc, it is similar to that estimated for the narrow collisionally ionized component.
The kinetic power of the narrow ionized components observed by XRISM is also similar, suggesting the inner ionized outflow could drive the outer infrared outflow.
\citet{Alonso2025} suggested the infrared ionized outflow could be driven by a jet interacting with the interstellar medium rather than the outflow.
Our results give the possibility that the AGN outflow could drive the infrared outflow.

\section{Conclusions}
In this study, we analyzed the broadband X-ray spectra (3–60 keV) of the nearby radio galaxy Centaurus A using joint XRISM/Resolve and NuSTAR observations. 
By performing a detailed spectral modeling with the SPEX code, we identified, for the first time, significant ionized emission components in the Fe-K band of Cen A. 
This marks a notable advancement in the study of circumnuclear environments in low-Eddington radio galaxies, where clear ionized features have rarely been reported.

The spectral modeling revealed three ionized emission components: one is broad and redshifted ($v_{\rm broad}=3100\pm500$ km s$^{-1}$, $v_{\rm shift}=4000\pm500$ km s$^{-1}$) and others are narrow redshifted and blueshifted ($v_{\rm broad}=400\pm100$ km s$^{-1}$, $v_{\rm shift}=2600\pm100$ km s$^{-1}$, and $v_{\rm broad}=550^{+110}_{-80}$ km s$^{-1}$, $v_{\rm shift}=-1500\pm100$ km s$^{-1}$), attributed to Fe {\sc xxv} and Fe {\sc xxvi} lines. 
The upper limit on the distance of the emitters is $1300~R_{\rm S}$ or $8\times10^3R_{\rm S}$, assuming a photoionized plasma. 
The collisionally ionized plasma model can reproduce the narrow line components, and in that case, the distance could be as large as 0.5 pc, but within the torus. 
We cannot distinguish whether the velocity is due to inflow, outflow, or rotation.

In addition, absorption lines at 7.1 keV and 10.6 keV were detected, corresponding to Fe {\sc xxv} absorption features. 
Modeling with the pion component suggests these are Fe {\sc xxv} lines blueshifted due to UFOs with outflow velocities of $\sim0.06c$ and $\sim0.43c$, respectively. 
These absorption components are characterized by moderate ionization parameters ($\log\xi\sim 2.9-3.3$) and relatively low column densities ($N_{\rm H}\sim10^{21}$ cm$^{-2}$). 
These results demonstrate the high potential of XRISM/Resolve for characterizing AGN outflows and emission features in the Fe-K band. 
Our findings establish a new benchmark for the spectroscopic study of low-luminosity radio galaxies, contributing to a broader understanding of AGN unification.

Based on our discussion and considering the synergy with the outer infrared outflow discovered by JWST, the broad component is interpreted as an inner region inflow and/or outflow with a wide range of velocities, and the UFOs might be a part of this inner outflow traveling at an extremely high velocity. The narrow components would be outflowing shock-heated plasma pushed by past AGN activity at around 0.5 pc, which could in turn drive the outer infrared outflow.

\begin{acknowledgments}
The authors thank the editors and the anonymous referee for their useful comments to improve the quality of this work.
This paper employs a list of Chandra datasets, obtained by the Chandra X-ray Observatory, contained in the Chandra Data Collection ~\dataset[DOI: 10.25574/cdc.514]{https://doi.org/10.25574/cdc.514}.
A reproduction package is available at Zenodo DOI: \href{https://zenodo.org/doi/10.5281/zenodo.17908257}{10.5281/zenodo.17908257}. This package includes data and scripts used to reproduce the fitting results and figures presented here.
This work was supported by JSPS KAKENHI Grant Number JP25KJ1868 (TK),
JP19K21884, JP20H01941, JP20H01947, JP20KK0071, JP23K20239, JP24K00672 (HN), JP23KJ0780 (TI), JP20H01946 (YU), JP21K13958 (MM),
JP24KJ1507 (YN),
JST SPRING Grant Number JPMJSP2132 (TK),
NASA under award number 80GSFC24M0006 (TY),
Canadian Space Agency grant 18XARMSTMA (LCG), 
Yamada Science Foundation (MM), and Tsinghua Dushi program (JM). 
We thank D. Rogantini, N. Kawakatsu, and M. Nomura for helpful discussions. 
\end{acknowledgments}

\bibliography{ref}{}
\bibliographystyle{aasjournalv7}

\appendix 
\section{Absorption line verification}
Whether the absorption lines are real is an important question and should be investigated to determine whether they are not pixel-independent, time-independent, or count-rate dependent. 
We verified the 7.1 keV and 10.6 keV absorption lines. 
We split the Resolve data by region, time frame, and flux level to verify the pixel dependence, time dependence, and flux dependence. 
When divided by time, various flux levels are included, and when divided by flux level, various observation times are included. 
The total count is $4.71\times10^{5}$ counts. 

\subsection{Region}
\begin{figure}[htb]
 \centering
 \includegraphics[width=0.7\hsize]{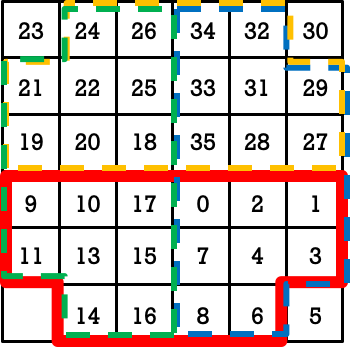}
 \caption{The XRISM/Resolve data was split into 16 pixels each for red, blue, yellow, and green regions. Referred from The XRISM Proposers’ Observatory Guide \footnote{\url{https://heasarc.gsfc.nasa.gov/docs/xrism/proposals/POG/Resolve.html}} (Fig. 5.1). }
 \label{region}
\end{figure}

\begin{figure}
 \centering
 \begin{minipage}[b]{0.45\hsize}
 \centering
 \includegraphics[width=\hsize]{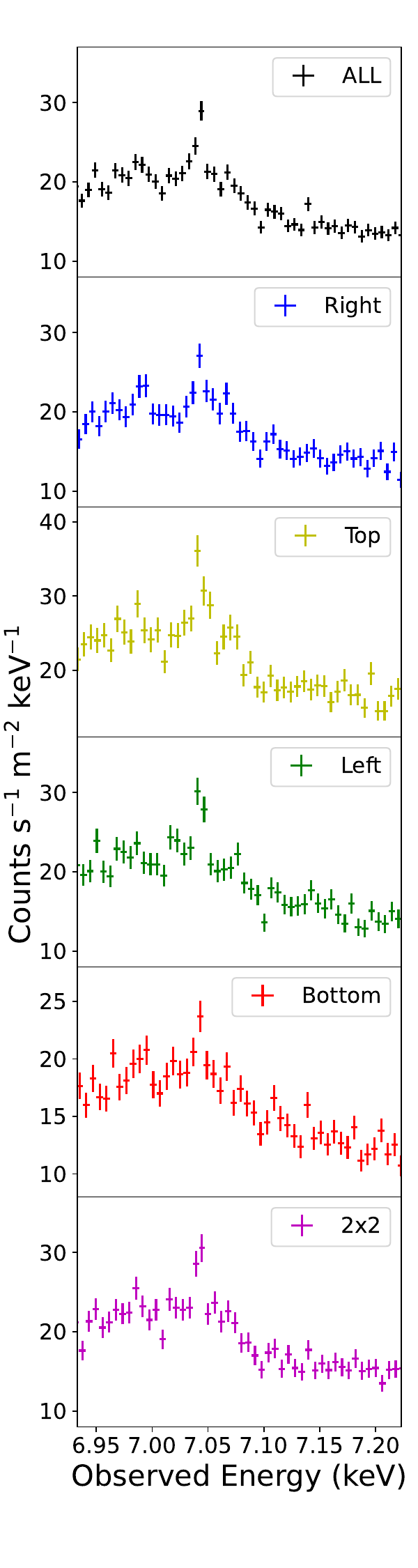}
 \subcaption{$6.8-7.4$ keV band}
 \label{RegionKb}
\end{minipage}
  \begin{minipage}[b]{0.45\hsize}
 \centering
 \includegraphics[width=\hsize]{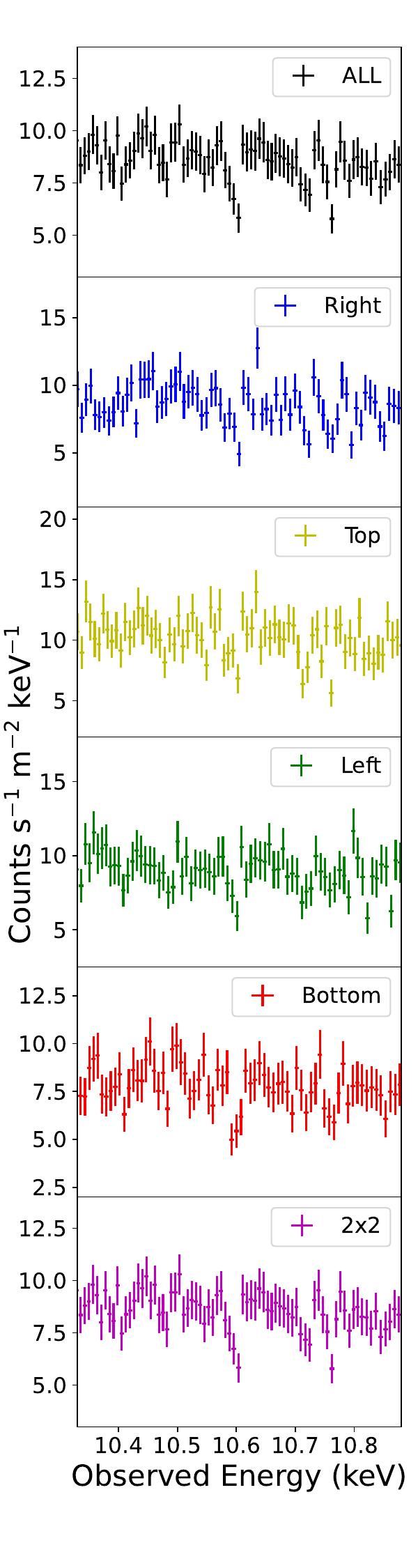}
  \subcaption{$10.4-10.95$ keV band}
   \label{Region10}
  \end{minipage}
 \caption{Region split Resolve spectra, wherein the black point, blue point, yellow point, green point, red point, and magenta point represent the total pixel, right region, top region, left region, bottom region, and central $2\times2$ pixel region spectrum, respectively. The spectrum is binned by a factor of 2 after the optimal binning for visual clarity.}
 \label{Regionspectra}
\end{figure}

First, we split the Resolve data into the regions (Fig. \ref{region}).
Each region contains 16 pixels. 
The counts for each region are listed in Table \ref{regionCounts}. 
From the region split spectra (Fig. \ref{Regionspectra}), the Fe-K$\beta$ emission line and the 10.6 keV absorption line are visible. 
The 7.1 keV is not easily seen, although an absorption line structure is required compared to the Fe-K edge.

\begin{table}[H]
 \caption{Counts for each region}
 \label{regionCounts}
 \centering
  \begin{threeparttable}
  \begin{tabular}{clccccll}
  \hline\noalign{\smallskip}
Region & Counts & Ratio \\
\noalign{\smallskip}\hline\noalign{\smallskip}
Bottom (red) & $2.32\times10^{5}$ & $0.49$ \\\noalign{\smallskip}
Right (blue) & $2.21\times10^{5}$ & $0.47$ \\\noalign{\smallskip}
Top (yellow) & $2.11\times10^{5}$ & $0.45$ \\\noalign{\smallskip}
Left (green) & $2.22\times10^{5}$ & $0.47$ \\\noalign{\smallskip}
Average & $2.22\times10^{5}$ & $0.47$ \\\noalign{\smallskip}
\hline
\end{tabular}
\begin{tablenotes}
\item
\end{tablenotes}
\end{threeparttable}
\end{table}

\subsection{Time frame}
\begin{figure}
 \centering
 \begin{minipage}[b]{0.45\hsize}
 \centering
 \includegraphics[width=\hsize]{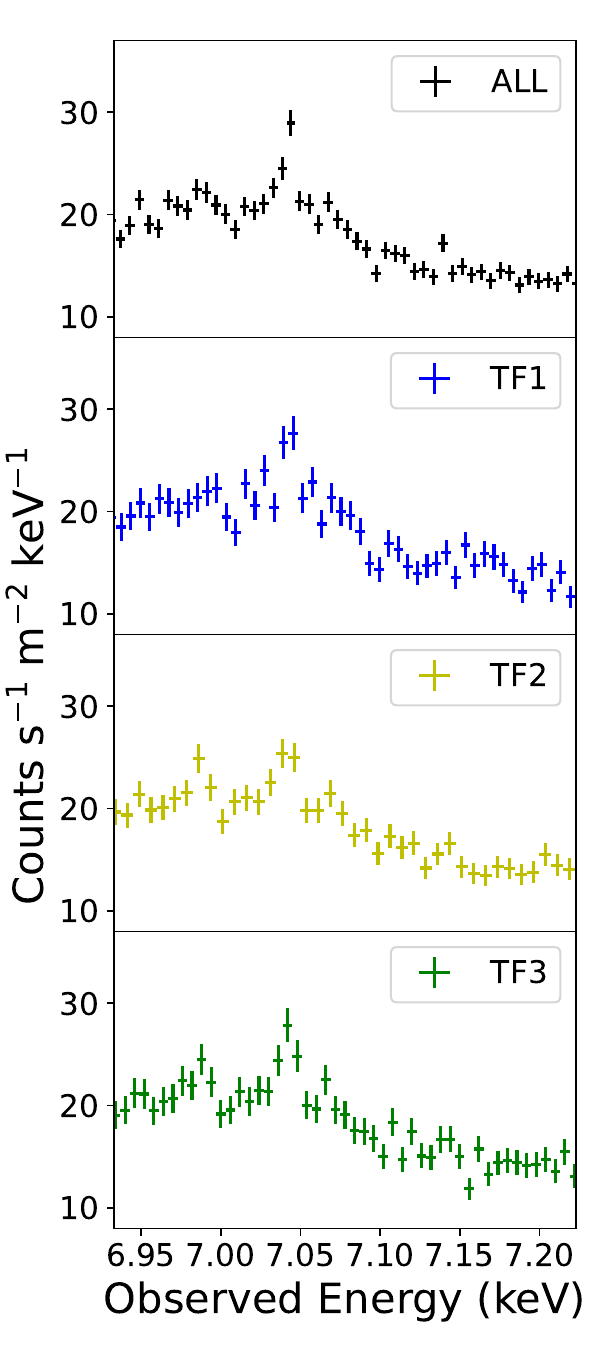}
 \subcaption{$6.8-7.4$ keV band}
 \label{TimeKb}
\end{minipage}
  \begin{minipage}[b]{0.45\hsize}
 \centering
 \includegraphics[width=\hsize]{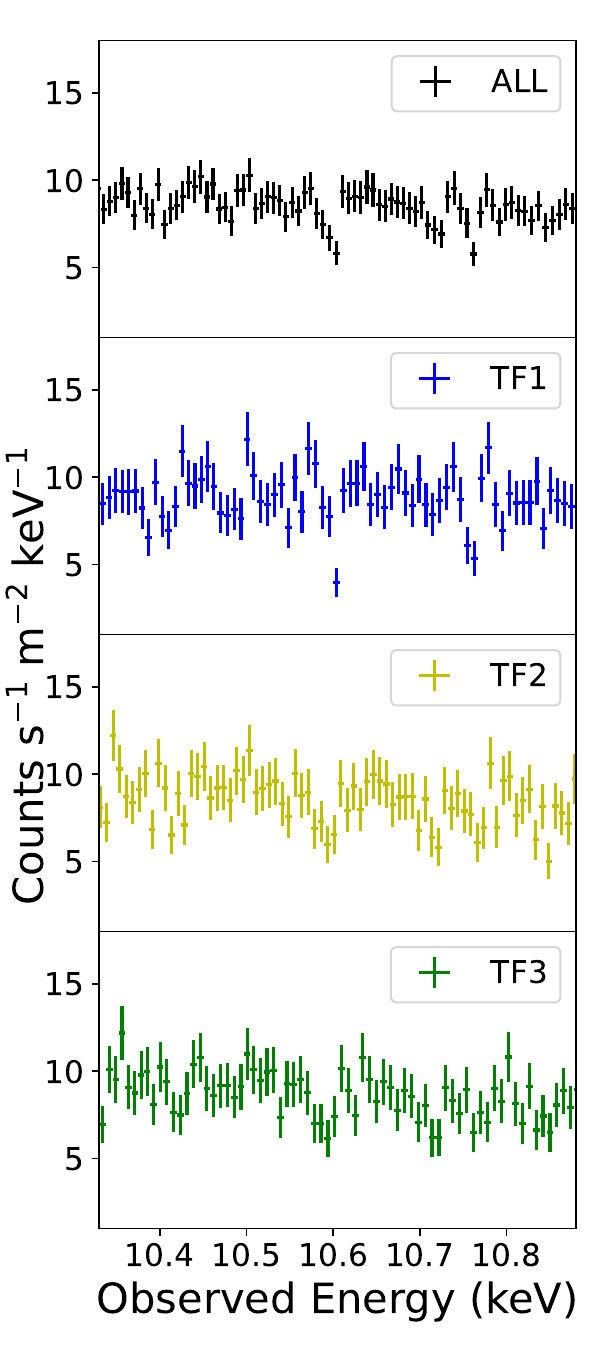}
  \subcaption{$10.4-10.95$ keV band}
   \label{Time10}
  \end{minipage}
 \caption{Time split Resolve spectra, wherein the black, blue, yellow, and green points represent the total time, timeframe 1, timeframe 2, and timeframe three spectra, respectively. The spectrum is binned by a factor of 2 after the optimal binning for visual clarity.}
 \label{Timespectra}
\end{figure}

The average count of the region split corresponds to $106.054$ ksec.
Therefore, we subtracted the data into the three timeframes to $106.054$ ksec. 
The first 106 ksec subtracted was Timeframe 1 (TF1), the middle was Timeframe 2 (TF2), and the last was Timeframe 3 (TF3). 
The line clarity from the time split spectra (Fig. \ref{Timespectra}) is similar to the region split spectrum, with the Fe-K$\beta$ emission line and 10.6 keV absorption lines seen. However, the 7.1 keV absorption line is difficult to see.

\subsection{Flux level}
\begin{figure}
 \centering
 \begin{minipage}[b]{0.45\hsize}
 \centering
 \includegraphics[width=\hsize]{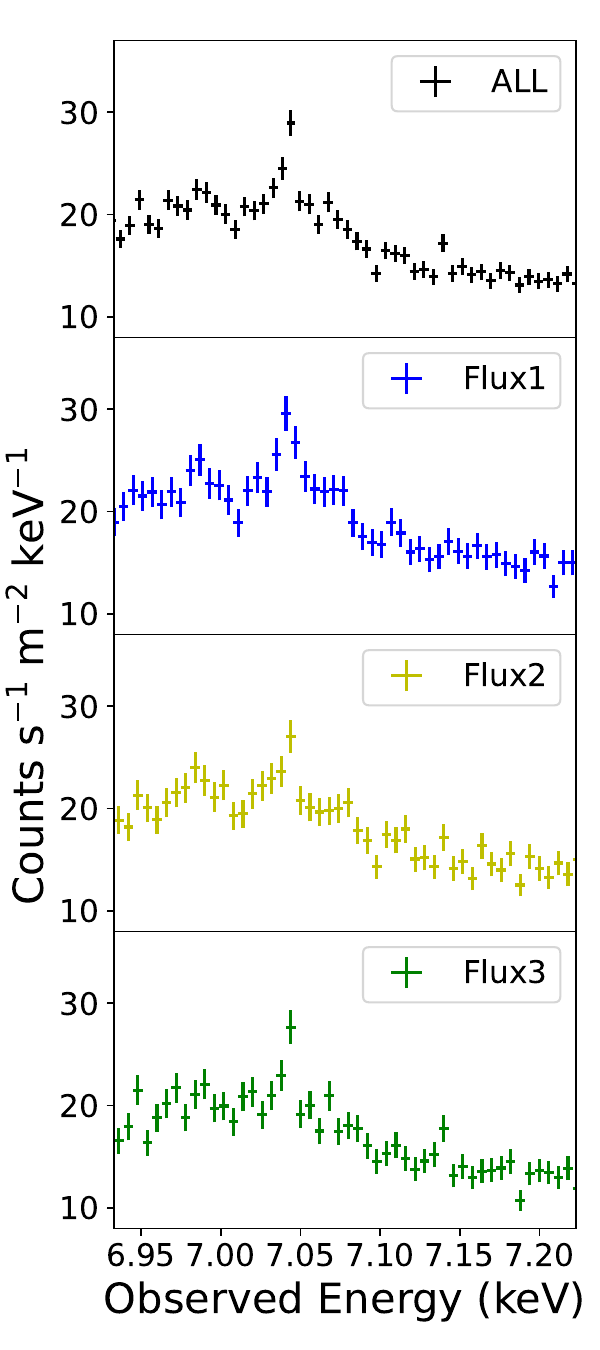}
 \subcaption{$6.8-7.4$ keV band}
 \label{FluxKb}
\end{minipage}
  \begin{minipage}[b]{0.45\hsize}
 \centering
 \includegraphics[width=\hsize]{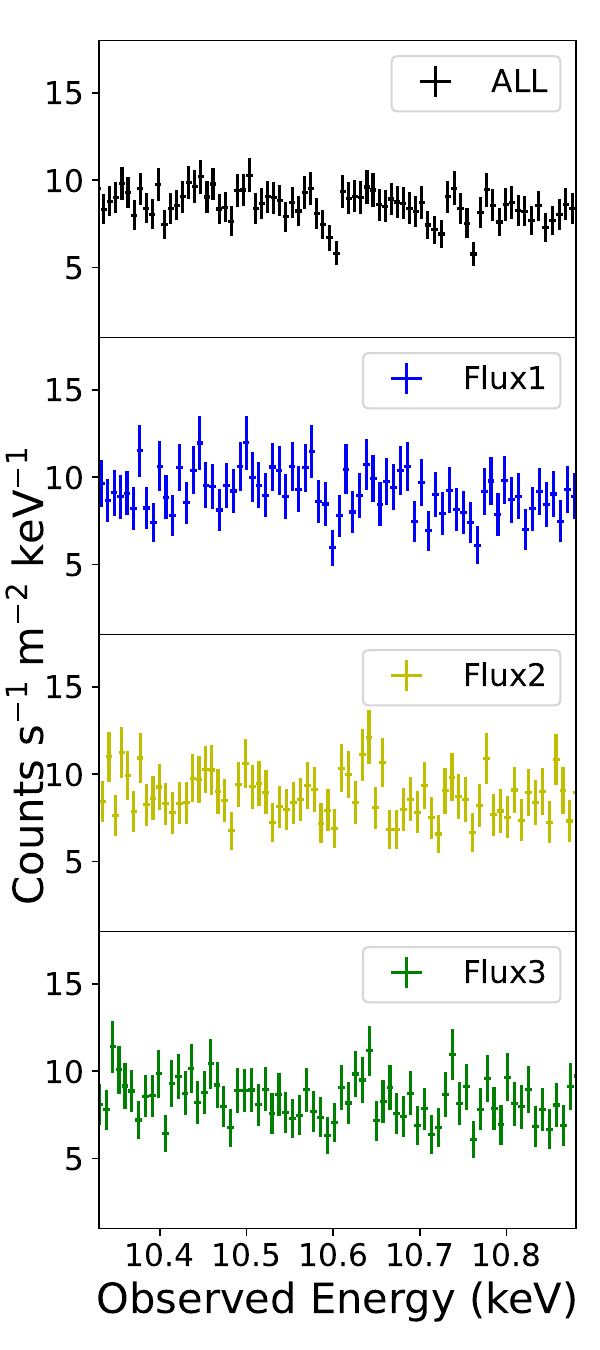}
  \subcaption{$10.4-10.95$ keV band}
   \label{Flux10}
  \end{minipage}
 \caption{Flux split Resolve spectra, wherein the black, blue, yellow, and green points represent the time, flux level 1, flux level 2, and flux level 3 spectra, respectively. The spectrum is binned by a factor of 2 after the optimal binning for visual clarity. }
 \label{Fluxspectra}
\end{figure}

We subtracted the data into the three flux levels (corresponding to $106.054$ ksec) as a timeframe. 
The data were sorted by count order and split into flux levels. 
Flux level 1 is a higher rate ($>1.98$ cts), Flux level 2 is a medium count rate ($1.30-2.06$ cts), and Flux level 3 is the lower rate ($<1.98$ cts).
From the flux split spectra (Fig. \ref{Fluxspectra}), the line clarity is similar to the region split spectrum, with the Fe-K$\beta$ emission line and 10.6 keV absorption lines seen. However, the 7.1 keV absorption line is difficult to see.

\subsection{Simulation}
\begin{figure}[htb]
 \centering
 \includegraphics[width=\hsize]{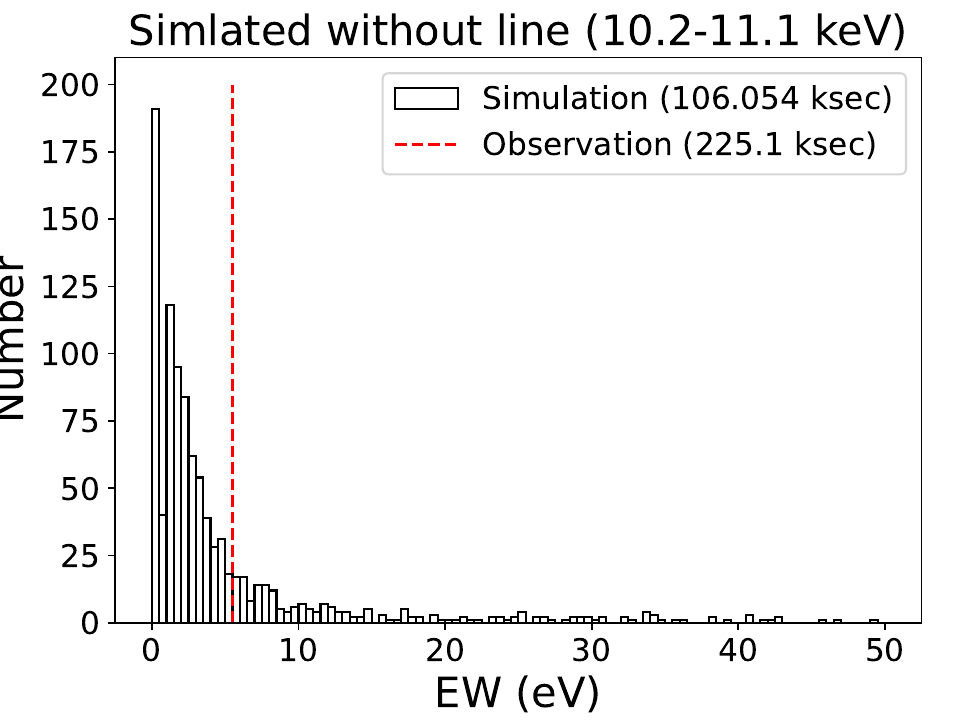}
 \caption{A Histogram of the EW. Simulated the $10.2-11.2$ keV Resolve spectrum and fitted with a power law (without the line). }
 \label{noLine}
\end{figure}

\begin{figure}[htbp]
 \centering
 \begin{minipage}[b]{\hsize}
 \centering
 \includegraphics[width=\hsize]{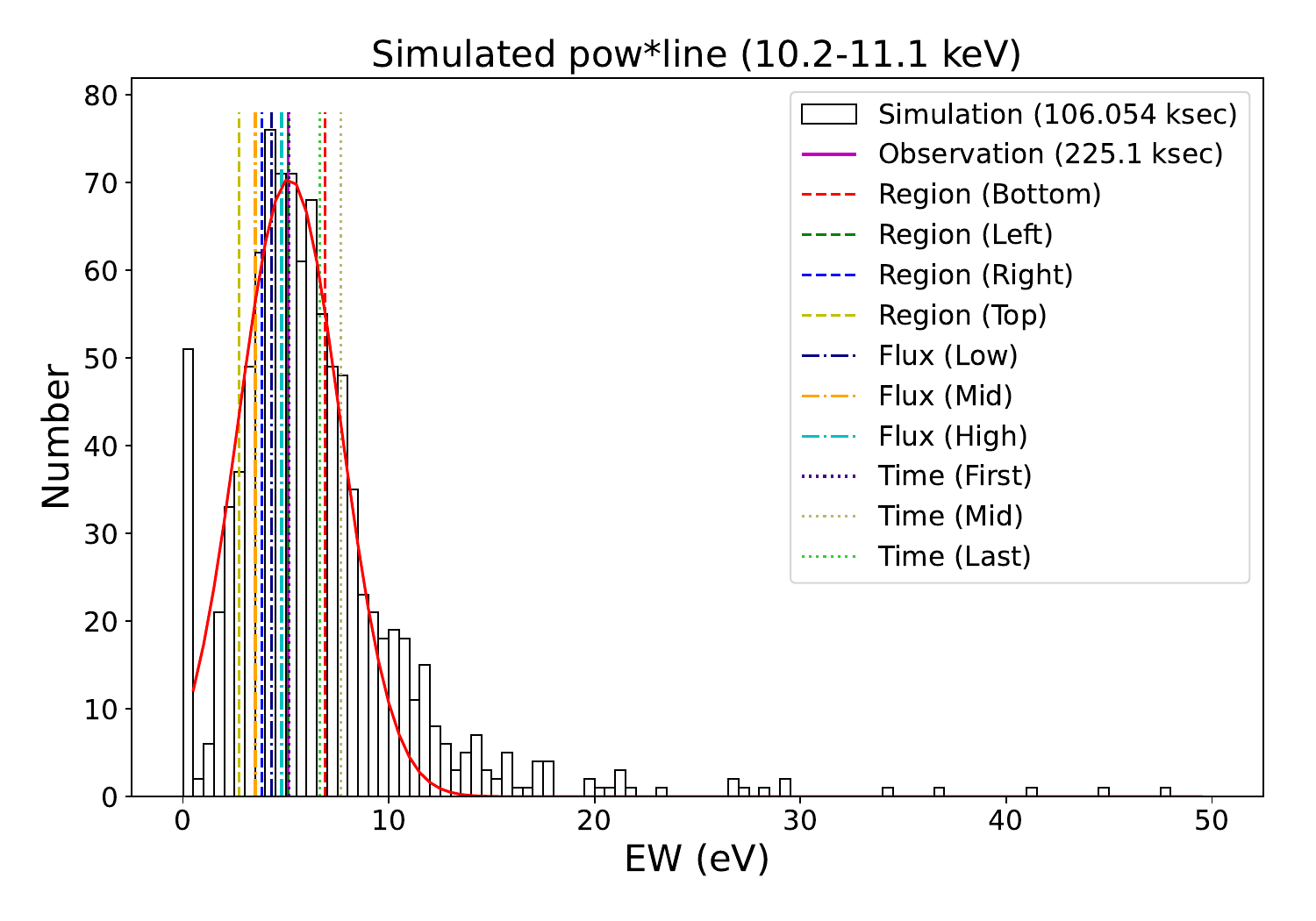}
\end{minipage}\\
  \begin{minipage}[b]{\hsize}
 \centering
 \includegraphics[width=\hsize]{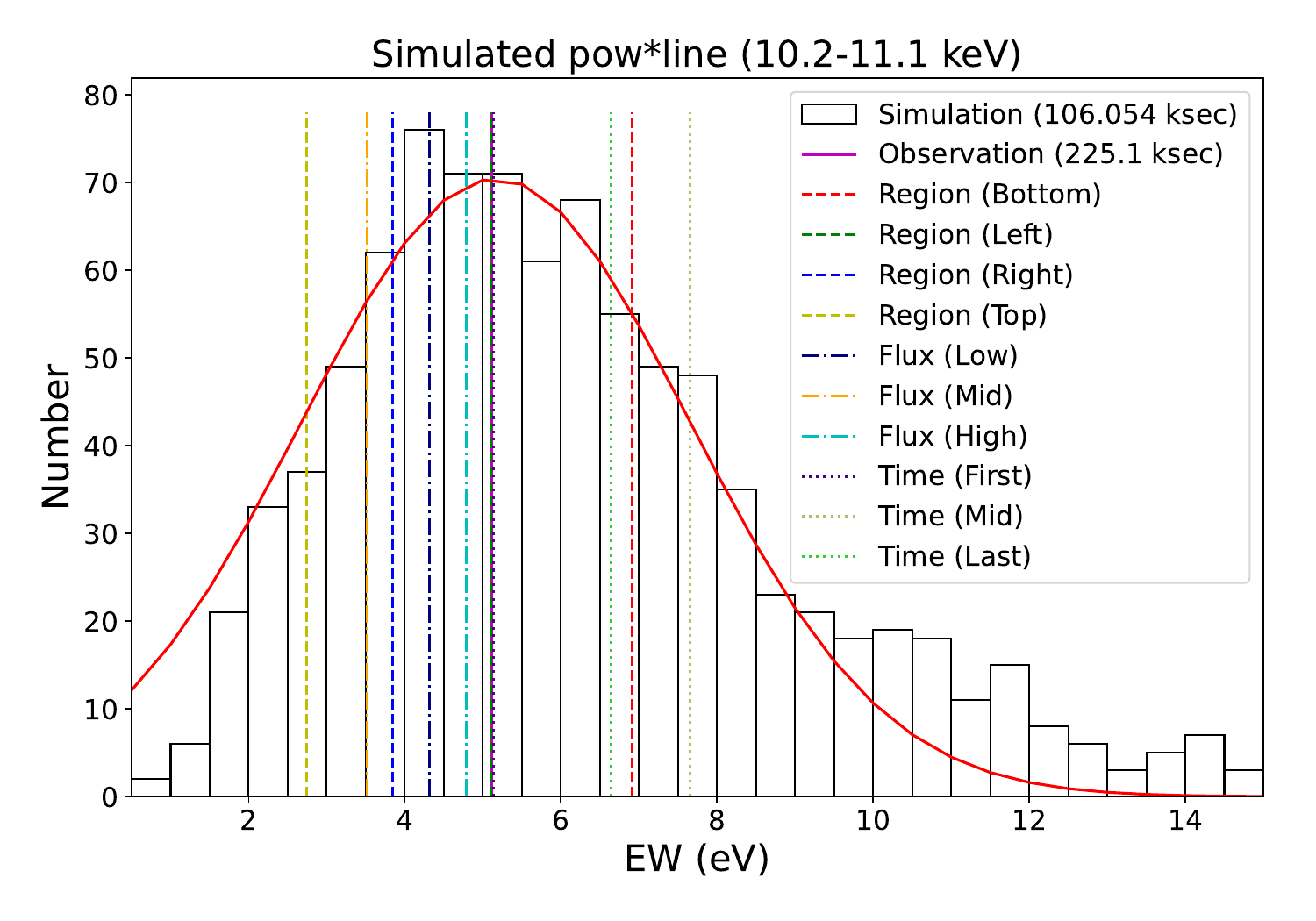}
  \end{minipage}
 \caption{A Histogram of the EW. Simulated the $10.2-11.2$ keV Resolve spectrum and fitted with a power law and a line model (Gaussian). A histogram was fitted with a Gaussian (red line, EW center is 5.17 eV, $\sigma$ is 2.5 eV). }
 \label{Sim}
\end{figure}

To investigate the significance of the 10.6 keV absorption line, we simulated the Resolve spectrum. 
For the 10.6 keV absorption line, we fitted the $10.2-11.2$ keV Resolve spectra with a power-law and line model (Gaussian) and obtained the $EW$ of the line ($EW=5.5$ eV). 
We simulated two cases (1,000 times for each case): the Resolve spectra, with and without (Fig. \ref{noLine} \& \ref{Sim}) an absorption line. 
The exposure time was set to 106.054 ksec as a subset of the Time frame and flux level. 
In addition, we fitted all the Region split spectra, the Flux split spectra, and the Time split spectra with a power-law and line model (Gaussian) and obtained the $EW$ of the line (Fig. \ref{Sim}). 
The histogram was fitted with a Gaussian; the $EW$ center is 5.17 eV, and $\sigma$ is 2.5 eV. 
The all case (The region, time, flux split spectra) is in the $1\sigma$. 
In the Fig. \ref{noLine} we assumed there is no absorption line in $10.2-11.2$ keV. 
We simulated the Resolve spectrum without the absorption line and fitted it with the line model. 
The histogram was fitted with a Gaussian; the $EW$ center is 1.52 eV, and $\sigma$ is 1.8 eV. 
Therefore, the 10.6 keV absorption line ($EW=5.5$ eV) is a $98.6\%$ level.

\section{Differences between H-like and He-like model curves in photo-ionization absorption}
\begin{figure}[htb]
 \centering
 \includegraphics[width=\hsize]{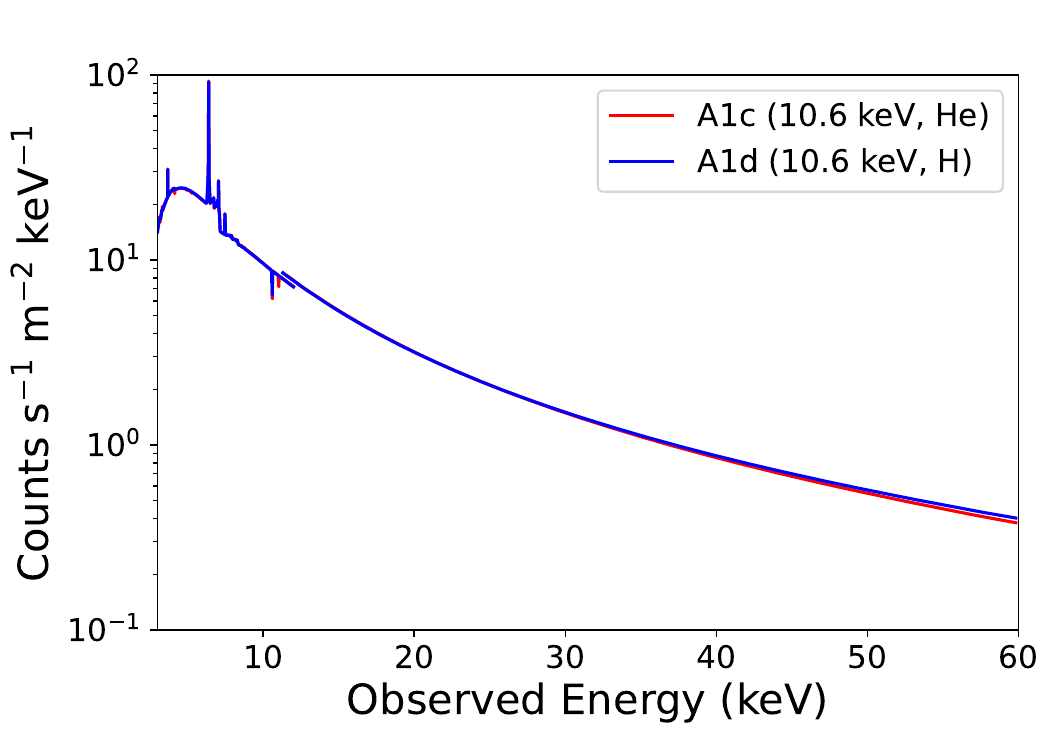}
 \caption{Differences between H-like and He-like model curves in photo-ionization absorption. The red line and the blue line represent the He-like Fe case and the H-like Fe case, respectively.  }
 \label{A1cdDiffModel}
\end{figure}

\section{UFOs with very fast outflow velocity}
\begin{table*}[htbp]
 \caption{The list of the UFOs with very fast outflow velocity ($\log v_{\rm out}$ [km s$^{-1}$]$\geq5$) in \citet{Yamada2024}. }
 \label{ufolist}
 \centering
 \begin{threeparttable}\small
  \begin{tabular}{lcccccccc}
\noalign{\smallskip}\hline\hline\noalign{\smallskip}
Object & Ra., Dec. & Redshift & Type & $\log v_{\rm out}$ & Reference \\
\noalign{\smallskip}\hline\noalign{\smallskip}
CXOCDFS J033260.0$-$274748 & 03h32m60.0s, $-$27d47m48s & 2.579 & QSO & $5.40$ & [1, 2] \\\noalign{\smallskip}
H 1413$+$117 & 14h15m46.2s, $+$11d29m44s & 2.560 & QSO & $5.30$ & [3] \\\noalign{\smallskip}
APM 08279$+$5255 & 08h31m41.7s, $+$52d45m18s & 3.910 & QSO & $5.02-5.25$ & [4-10] \\\noalign{\smallskip}
HS 1700$+$6416 & 17h01m00.6s, $+$64d12m09s & 2.735 & QSO & $5.06-5.24$ & [10, 11] \\\noalign{\smallskip}
SDSS J1029$+$2623 & 10h29m11.9s, $+$26d23m38s & 2.197 & QSO & $5.24$ & [10] \\\noalign{\smallskip}
SDSS J1128$+$2402 & 11h28m18.5, $+$24d02m18s & 1.608 & QSO & $5.24$ & [10] \\\noalign{\smallskip}
SDSS J0921$+$2854 & 09h21m15.5s, $+$28d54m44s & 1.410 & QSO & $5.09-5.20$ & [10] \\\noalign{\smallskip}
NGC 4151 & 12h10m32.6s, $+$39d24m21s & 0.0033 & 1.5 & $5.24$ & [12] \\\noalign{\smallskip}
PDS 456 & 17h28m19.8s, $-$14d15m56s & 0.1840 & 1/QSO & $5.03-5.16$ & [13-15] \\\noalign{\smallskip}
SDSS J1442$+$4055 & 14h42m54.8s, $+$40d55m36s & 2.593 & QSO & $5.15$ & [10] \\\noalign{\smallskip}
PKS 1549$-$79 & 15h56m58.9s, $-$79d14m04s & 0.1522 & NLS1 & $5.11$ & [16] \\\noalign{\smallskip}
HS 0810$+$2554 & 08h13m31.3s, $+$25d45m03s & 1.51 & QSO & $5.09-5.11$ & [10, 17-18] \\\noalign{\smallskip}
Mrk 231 & 12h56m14.2s, $+$56d52m25s & 0.0422 & 1.0 & $5.10$ & [19] \\\noalign{\smallskip}
PG 1211$+$143 & 12h14m17.7s, $+$14d03m13s & 0.0809 & NLS1 & $5.10$ & [20] \\\noalign{\smallskip}
NGC 2992 & 09h45m42.1, $-$14d19m35s & 0.0077 & 1.9 & $5.02-5.09$ & [21] \\\noalign{\smallskip}
IRAS 00521$-$7054 & 00h53m56.2s, $-$70d38m04s & 0.0689 & 2.0 & $5.08$ & [22] \\\noalign{\smallskip}
PG 1115$+$080 & 11h18m16.9s, $+$07d45m59s & 1.733 & QSO & $5.05$ & [3, 10, 23] \\\noalign{\smallskip}
Ark 564 & 22h42m39.3s, $+$29d43m31s & 0.0240 & NLS1 & $5.03$ & [24] \\\noalign{\smallskip}
MCG-03-58-007 & 22h49m37.1s, $-$19d16m26 & 0.0323 & 2.0 & $5.01-5.03$ & [25-28] \\\noalign{\smallskip}
SDSS J1353$+$1138 & 13h53m06.3s, $+$11d38m05s & 1.627 & QSO & $5.01$ & [10] \\\noalign{\smallskip}
\hline
\end{tabular}
\begin{tablenotes}
\item 
$\log v_{\rm out}$: Logarithm of the outflow velocity in unit km s$^{-1}$. References; (1) \citet{Wang2005}, (2) \citet{Zheng2008}, (3) \citet{Chartas2007}, (4) \citet{Chartas2002}, (5) \citet{Chartas2009}, (6) \citet{Saez2011}, (7) \citet{Gofford2013}, (8) \citet{Gofford2015}, (9) \citet{Hagino2017}, (10) \citet{Chartas2021}, (11) \citet{Lanzuisi2012}, (12) \citet{Zhou2019}, (13) \citet{Reeves2018}, (14) \citet{Boissay2019}, (15) \citet{Reeves2020}, (16) \citet{Tombesi2014}, (17) \citet{Chartas2014}, (18) \citet{Chartas2016}, (19) \citet{Mizumoto2019b}, (20) \citet{Reeves2005}, (21) \citet{Luminari2023}, (22) \citet{Walton2019}, (23) \citet{Chartas2003}, (24) \citet{Kara2017}, (25) \citet{Braito2018}, (26) \citet{Matzeu2019}, (27) \citet{Braito2021}, (28) \citet{Braito2022}
\end{tablenotes}
\end{threeparttable}
\end{table*}

\end{document}